\documentclass[useAMS,usenatbib,aps,floatdix,nofootinbib]{mn2e}

\usepackage{graphicx}

\usepackage{amssymb,amsmath}

\usepackage{psfrag}
\usepackage{graphicx}

\usepackage{color}
\usepackage{ulem}
\usepackage{xspace}
\usepackage{caption}
\usepackage{subcaption}
\usepackage[hidelinks=true]{hyperref}
\usepackage{enumitem}
\usepackage{comment}
\usepackage{lineno}

\newcommand{\f}{\frac}

\newcommand{\be}{\begin{equation}}      
\newcommand{\ee}{\end{equation}}      
      
\newcommand{\bef}{\begin{figure}}      
\newcommand{\eef}{\end{figure}}      
\newcommand{\bea}{\begin{eqnarray}}    
\newcommand{\eea}{\end{eqnarray}}      
  
\def\spose#1{\hbox to 0pt{#1\hss}}
\def\ltapprox{\mathrel{\spose{\lower 3pt\hbox{$\mathchar"218$}}
\raise 2.0pt\hbox{$\mathchar"13C$}}}
\def\gtapprox{\mathrel{\spose{\lower 3pt\hbox{$\mathchar"218$}}
\raise 2.0pt\hbox{$\mathchar"13E$}}}
\def\inapprox{\mathrel{\spose{\lower 3pt\hbox{$\mathchar"218$}}
\raise 2.0pt\hbox{$\mathchar"232$}}}

\def\bse{\begin{subequations}}
\def\ese{\end{subequations}}

\def\lsim{\raise 0.4ex\hbox{$<$}\kern -0.8em\lower 0.62ex\hbox{$\sim$}} 
\def\gsim{\raise 0.4ex\hbox{$>$}\kern -0.7em\lower 0.62ex\hbox{$\sim$}}

\def\f0N{f_0^{(N)}}
\def\bec{\begin{center}}
\def\eec{\end{center}}

\newcommand{\Abacus}{\textsc{Abacus}\xspace}
\newcommand{\Rockstar}{\textsc{Rockstar}\xspace}
\newcommand{\FOF}{\textsc{fof}\xspace}

\title[Testing Dark Matter Halo Properties  ]
{Testing dark matter halo properties using self-similarity}
\author[M. Leroy et al.]
{M. Leroy${^{1}}$, L. Garrison${^{2,3}}$,
D. Eisenstein${^{3}}$, M. Joyce${^{4}}$\thanks{Corresponding author email: joyce@lpnhe.in2p3.fr} and S. Maleubre${^{4}}$\\ \\
$^1$\'Ecole Normale Sup\'erieure Paris-Saclay, 61 Avenue du Président Wilson, 94230 Cachan, France \\ 
$^2$Harvard-Smithsonian Center for Astrophysics, 60 Garden St., MS-10, Cambridge, MA 02138, USA\\
$^3$Center for Computational Astrophysics, Flatiron Institute, 162 Fifth Ave., New York, NY 10010, USA\\
$^4$Laboratoire de Physique Nucl\'eaire et de Hautes \'Energies, UPMC IN2P3 CNRS UMR 7585, \\ \,\,Sorbonne Universit\'e, 4, place Jussieu, 75252 Paris Cedex 05, France \\
 }

\begin{document}

\date{\today}

\maketitle

\begin{abstract}
We use self-similarity in $N$-body simulations of scale-free models to test for resolution dependence in the mass function and two-point correlation functions of dark matter halos. We use 1024$^3$ particle simulations performed with \Abacus, and compare results obtained with two halo finders:  friends-of-friends (\FOF) and \Rockstar. 
The \FOF mass functions show a systematic deviation from self-similarity which is explained by resolution dependence of the \FOF mass assignment previously reported in the literature.
Weak evidence for convergence is observed only starting from halos of several thousand particles, and mass functions are overestimated by at least as much as $20-25\%$ for halos of 50 particles. The mass function of the default \Rockstar halo catalog (with bound virial spherical overdensity mass), on the other hand, shows good convergence from of order 50 to 100 particles per halo, with no detectable evidence at the 
few percent level of any systematic dependence for larger particle number. 
Tests show that the mass unbinding procedure in \Rockstar is the key factor in obtaining this much improved resolution. Applying the same analysis
to the halo-halo two point correlation function, we find again strong
evidence for convergence only for \Rockstar halos, at separations sufficiently large so that halos do not overlap. At these separations 
we can exclude dependence on resolution at the $5-10\%$  level
once halos have of order 50 to 100 particles. At smaller separations 
results are not converged even at significantly larger particle 
number, and bigger simulations would be required to establish 
the resolution required for convergence. 

\end{abstract}


\section{Introduction}

The approximate description of the cosmological matter field in terms of a decomposition in ``halos'' has become a central tool of cosmological structure formation (see e.g. \cite{cooray+sheth_2002}). With the advent of ever more precise observational data, the issue of the precision of this description has become an important practical one. Indeed as theoretical predictions for many observables (e.g. galaxy-galaxy correlation functions) are obtained using constructions based on the halo decomposition, their precision relies on that of the latter. The issue of the precision of relevant halo properties, which are obtained uniquely from $N$-body simulations, is particularly complex as it combines two distinct issues: that of the precision with which mass distribution in the $N$ body simulation represents the physical limit and that of the halo definition and extraction. Halos are not objects with a unique definition and they are defined in practice by the algorithm adopted to extract them from the $N$-body simulation. Numerous different ways have been proposed and exploited (see \cite{knebe_et_al_2013} for a review). In this article we apply a test of the accuracy with which basic properties of halos, as determined by two different widely used algorithms, can be obtained from cosmological simulations. Specifically we focus on the resolution limits due to the finite particle density used to sample the density field in the $N$-body method. 

The analysis follows very closely that we have reported in a recent paper 
(\cite{JEG_Paper1}, hereafter P1) in which we have shown how very precise constraints on the convergence to physical values of the two point correlation function (2PCF) of the full matter field can be obtained by studying the {\it deviations} from self-similarity in its evolution in simulations of a scale-free model. Indeed  if such deviations are observed they necessarily imply a dependence of the result on unphysical parameters. In the direct study of the matter field considered in P1, such parameters can only be those introduced by the $N$-body method, i.e., the box size $L$, initial grid spacing $\Lambda$, force softening $\varepsilon$ and the parameters characterising initial conditions (IC) (as well as the numerical parameters controlling the accuracy of the $N$-body integration). The halos and their properties as defined by the halo extraction will also necessarily depend at some level  on at least some of these unphysical parameters --- in particular $\Lambda$ which determines the particle density --- and may, depending on the algorithm, introduce other such parameters. Thus by testing for the self-similarity of halo statistics we can simultaneously test for both accuracy limits arising from the underlying field on which the halos are sampled and for accuracy limits arising from the method used to extract the halos. 

As the analysis we employ here is completely parallel in its essentials to that presented in detail in P1, we limit ourselves here to a very brief recapitulation of the essential steps with an emphasis on the points which are specific to the halo statistics we consider here. As noted in P1, the self-similarity of scale-free models has been widely recognised since the early development of $N$-body simulation in cosmology as an instrument to
check the reliability of simulation results (e.g. \cite{efstathiou1988gravitational,colombi1995self}), and such models have 
been used quite extensively in  the study of halo properties (e.g. \cite{Navarro_etal_1997, cole1996structure, knollmann_etal_2008, elahi_etal_2009, diemer+kravtsov_2015, ludlow+angulo_2017, diemer+joyce_2019}), we also refer to P1 for further references to previous literature. The required self-similarity of these models has, however, not been exploited in the way we do here to extract quantitative constraints on resolution. 

In scale-free simulations, the initial power spectrum of fluctuations is a power law $P(k) \propto k^{n}$, where $n$ is a constant and the expansion law is that Einstein de Sitter (i.e. $a(t) \propto t^{2/3}$). One can then infer that, if there is no dependence on any other length or time scales introduced, any statistic must be ``self-similar'' , i.e. invariant in time if expressed in terms of suitably rescaled space (or mass) variables. 
For any statistic written as a dimensionless function
$F(x_1,x_2,\cdots ;a)$ of the quantities $x_1,x_2 \cdots$ (e.g. separations, angles, masses) it depends on, self-similarity can be expressed simply as
\begin{equation}
F(x_1,x_2\cdots ;a) = F_0(x_1/X1_{\mathrm{NL}}(a),x_2/X2_{\mathrm{NL}}(a),\cdots),
\label{ss-scaling}
\end{equation}
where each  of the $Xi_{\mathrm{NL}}(a)$ is the temporal dependence of 
any quantity with the dimensions of $x_i$ inferred from self-similar 
scaling. The characteristic length scale $R_{NL}$ can be 
defined by
\begin{equation}
\sigma_{\mathrm{lin}}^{2}(R_{\mathrm{NL}},a) = 1,
\label{rnl_def}
\end{equation}
where $\sigma^2_{\rm lin}(R,a)$ is the linear theory normalized 
variance of mass in a sphere of radius $R$ at scale-factor $a$.
Defining (following the notation of P1) 
$\sigma_i=\sigma_{\rm lin}(\Lambda,a_i)$ where $a_i$ is the 
value of the scale factor at the start of the simulation,
we can infer that
\begin{equation}
R_{\mathrm{NL}}(a) = \Lambda \left( \frac{a}{a_i} \sigma_i \right)^{2/(n+3)}\,.
\end{equation}
The characteristic mass scale can then be defined as 
\begin{equation}
M_{\mathrm{NL}}(a) \equiv \frac{4\pi}{3} \bar{\rho} R^3_{\mathrm{NL}}(a)=
\frac{4\pi}{3} m_P \left( \frac{a}{a_i} \sigma_i \right)^{6/(n+3)}\,,
\end{equation}
where $\bar{\rho}$ is the mean (comoving) mass density and $m_P$ is the mass of a simulation particle. 

The two statistics which we consider here are the halo mass function (HMF) and the halo-halo correlation functions. Defining $n(M,a)$ canonically as the number of halos per unit mass interval and per unit comoving volume, we can express conveniently the self-similarity of the HMF as 
\begin{equation}
\frac{ M_{\mathrm{NL}(a)}^2 }{\bar{\rho}} n(M,a)= h_0 (M/M_{\mathrm{NL}}(a)),
\label{n-ss}
\end{equation}
where $h_0$ is a function.
The 2PCF $\xi_{HH} (r,M)$ of halo centres is a function of separation $r$
and of the halo mass $M$. As it is dimensionless, its self-similarity is simply expressed as 
\begin{equation}
\xi_{HH} (r,M,a)=\xi_{HH,0} (r/R_{\mathrm{NL}}(a),M/M_{\mathrm{NL}}(a)).
\label{xi-ss}
\end{equation}
It is the observed deviations from these scalings that we use here to infer the limitations on accuracy arising in particular from finite particle density.

\section{Simulations}
As described in P1, we have performed scale-free simulations using the \Abacus cosmological N-body code. 
Further background on the \Abacus code can be found in  \cite{2018ApJS..236...43G} and \cite{garrison_et_al_2019}, and further detail will be provided 
in forthcoming publications (Garrison et al.~(in prep.), Pinto et al.~(in prep.). 
We consider here results for the reference simulation in P1: this is a simulation with $N=1024^3$ particles of  an $n=-2$ scale-free model with force softening length $\varepsilon = \Lambda/30$. The Gaussian initial conditions are specified by 
$\sigma_i=0.03$, and by a choice of a scale factor fixing  corrections applied for discreteness effects at early times.
As discussed in P1 we have chosen values of the  numerical parameters controlling the $N$-body integration --- in particular the value of time-stepping parameter, $\eta=0.15$ --- for which numerous tests have shown that the statistics we study here are converged (with respect to these parameters) to a precision well below the percent  level. We have, in particular, in P1 reported a comparison with simulations with different values of $\eta$ which confirm this conclusion.

Most of our results will be given in terms of dimensionless quantities.
As we will focus primarily on resolution effects associated with the finite particle density, the associated bounds on mass will be given in units 
of the particle mass. 
As time variable we use, as in P1, $\log_2(a/a_0)$, where 
the reference $a_0$ is defined by
\begin{equation}
\sigma_{\rm lin} (\Lambda, a_0)= 0.56 \,, 
\label{a0-ref}
\end{equation}
which is a simple estimate for the time at which the first non-linear structures appear in the simulation ($0.56 \approx \delta_c/3$ where $\delta_c$ is the estimated threshold linear density fluctuation for virialization). We take our first output at 
$a=a_0$, and as in P1, subsequent outputs at scale-factors
$\{a_1,a_2 \cdots a_P\}$ with equal logarithmic spacing 
chosen as
\begin{equation}
\frac{M_{\mathrm{NL} }(a_{i+1})}{M_{\mathrm{NL} }(a_i)}=\sqrt{2},
\label{eq-snapshot-spacing}
\end{equation}
i.e. for every two intervals the theoretical non-linear mass scale 
grows by a factor of $2$. The last output corresponds to $P=37$,
at which time the largest halos contain of order a 
million particles\footnote{Although we do not use redshift $z$ 
here, it may be instructive to note that if we define it so that
$z=0$ at our final time, the starting red-shift of 
our simulation is $z_i \approx  157$, while $a=a_0$ corresponds to $z \approx 7.5$.}.

\subsection{Halo Extraction}

For each of our 38 outputs, we have constructed two halo catalogs using the Friends of Friends (\FOF) algorithm (see \cite{knebe_et_al_2013} for references) and the \Rockstar code \citep{2013ApJ...762..109B}. 

\FOF uses a single parameter $b$ (linking length) to link particles separated by a distance smaller than $b\Lambda$. The most common choice for its value is $b=0.2$ which corresponds to a theoretical threshold density of $\sim 80$ times the mean matter density \citep{2011ApJS..195....4M}.
Here we use $b=0.138$, corresponding to a critical density of approximately $240$ times the mean matter density. Our generated catalog retains only halos of at least 25 particles. 

The  publicly available \Rockstar (Robust Overdensity Calculation using K-Space Topologically Adaptive Refinement, \cite{2013ApJ...762..109B} ) code uses a much more sophisticated algorithm, involving an initial identification of \FOF groups in three dimensional space which are then successively refined using a \FOF algorithm in phase space. This  leads to a nested hierarchy of halos which are classified as `parent halos' or `subhalos' (with these themselves further refined as 'child' or successive generations). The mass associated with the halos is defined using various spherical overdensity(SO) criteria (which can be input or taken at default values), and, further, is refined by an identification of mass as gravitationally bound or unbound. 

 Of the numerous outputs provided by \Rockstar we restrict ourselves here to using {\it the SO mass corresponding to the virial radius } i.e., the mass within a spherical region about the halo center in which the mean density is greater than $178$ times the background density (in an EdS universe). In most of our analysis below we {\it include all halos and consider only the gravitationally bound mass}. In Section \ref{Comparison-catalogs} we consider how our analysis is modified 
by including also the gravitationally unbound mass, and also explore the `parent halo' and `subhalo' catalogs separately. 
We have run with the default parameters of the  \Rockstar code, except that we have set the  parameter fixing the minimal size of the initial \FOF groups taken as seeds to 25. The output catalogs, which contain halos down to two particles, are  expected thus to be incomplete when halos have less than 25 particles. 

\section{Halo Mass Function}

\begin{figure}
\centering\resizebox{8cm}{!}{\includegraphics[]{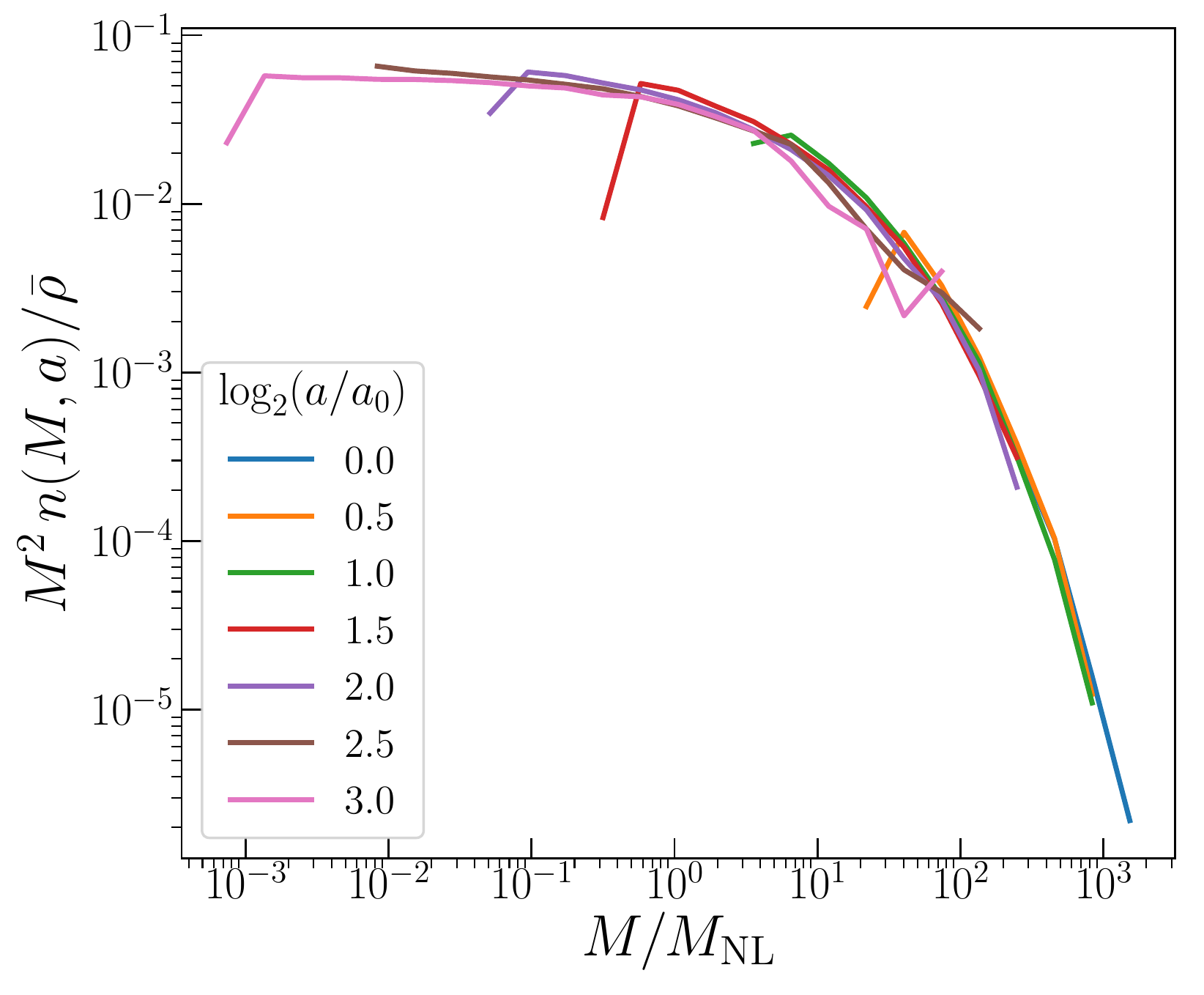}}
\caption{Rescaled mass function of \FOF halos, multiplied by $(M/M_{\text{NL}})^2$ for clarity. Approximate self-similarity, corresponding to superposition of the curves, is observed between lower and upper cut-offs determined by the minimal and maximal mass resolved.}
\label{fig:FOF_HMF}
\end{figure} 

\begin{figure}
\centering\resizebox{8cm}{!}{\includegraphics[]{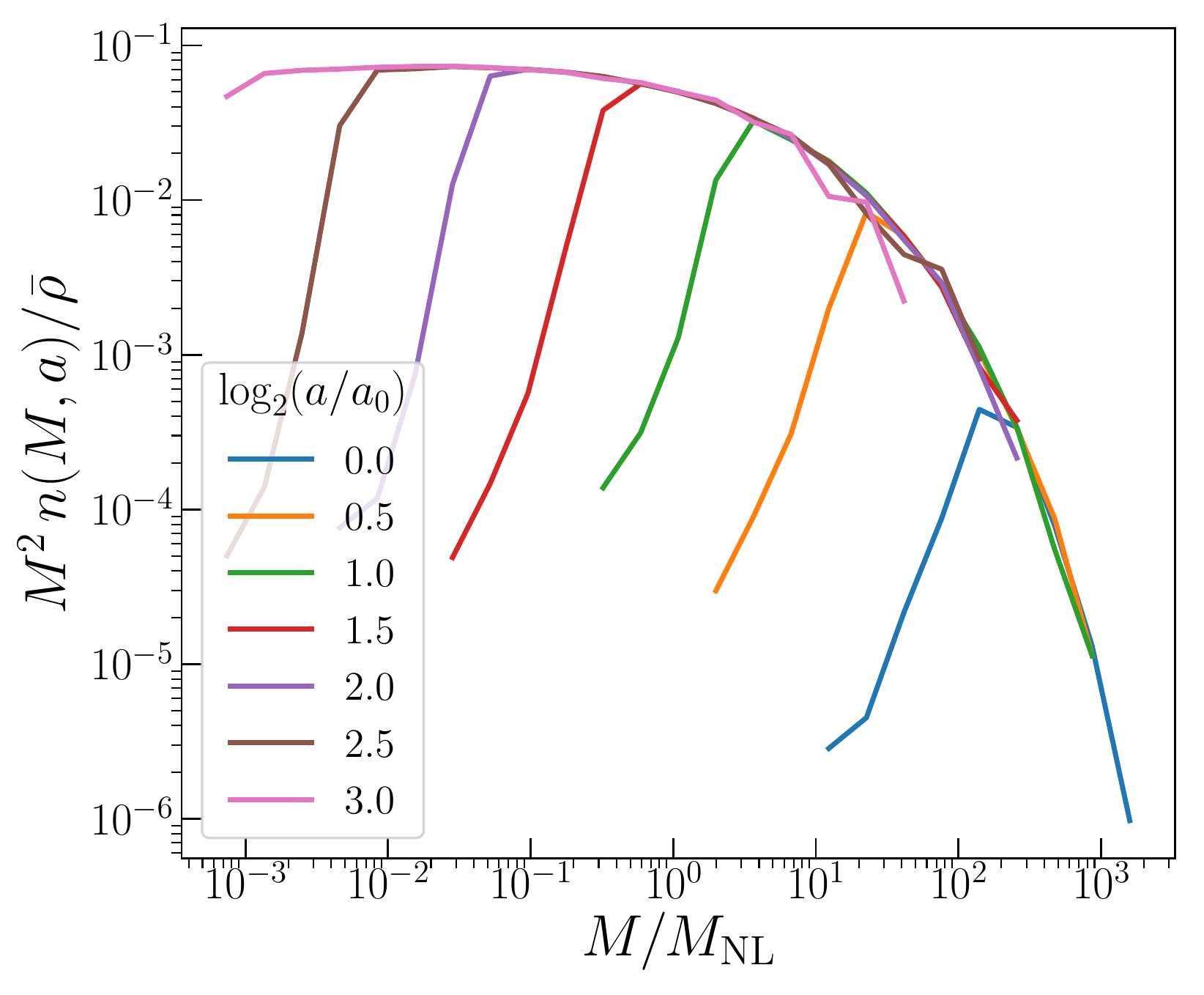}}
\caption{Rescaled mass function of \Rockstar halos, multiplied by $(M/M_{\text{NL}})^2$. As in the previous figure we observe superposition (and thus self-similarity) to a good approximation. The behaviour below the lower cut-off to self-similarity is different to
that for \FOF simply because our \Rockstar catalog includes halos down to two particles.}
\label{fig:ROCKSTAR_HMF}
\end{figure} 

As we have discussed, if the HMF of extracted halos corresponds to its physical (continuum) limit it should be self-similar. This means that
the measured value of $M_{\mathrm{NL}}^2 n(M,a)$, where $n(M,a)$ is the usual HMF, should be invariant as a function of time, when plotted 
as a function $M/M_{\mathrm{NL}}$.   Figures \ref {fig:FOF_HMF} and \ref{fig:ROCKSTAR_HMF} show the corresponding
plots for the \FOF and \Rockstar halos, respectively. For convenience of comparison, we have multiplied $M_{\mathrm{NL}}^2 n(M,a)/\bar{\rho}$
by $(M/M_{\mathrm{NL}})^2$ to obtain a plot for the fraction of mass in halos  per logarithmic interval of mass. For clarity we show in each figure only a subsample of
7 of our 38 available outputs. As specified above, the mass used for the  \Rockstar analysis is the gravitationally bound SO virial mass and all halos are included in the catalog.

\begin{figure*}
    \centering
    \begin{subfigure}[b]{0.35\textwidth}
    \includegraphics[width=0.9\textwidth,height=0.6375\linewidth]{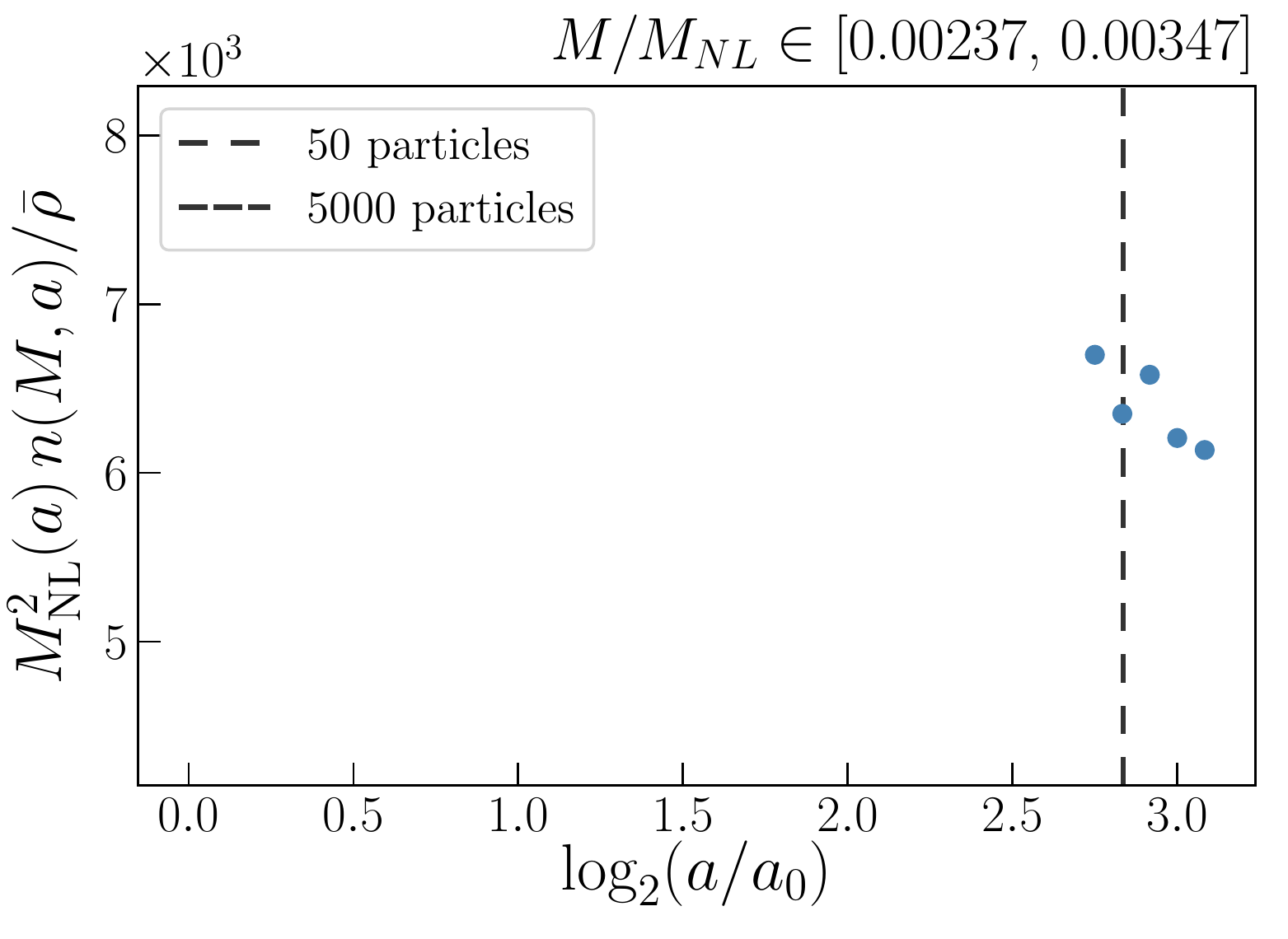}
    \end{subfigure}
    \hspace*{-0.7cm}    \begin{subfigure}[b]{0.35\textwidth}
    \includegraphics[width=0.9\textwidth,height=0.6375\linewidth]{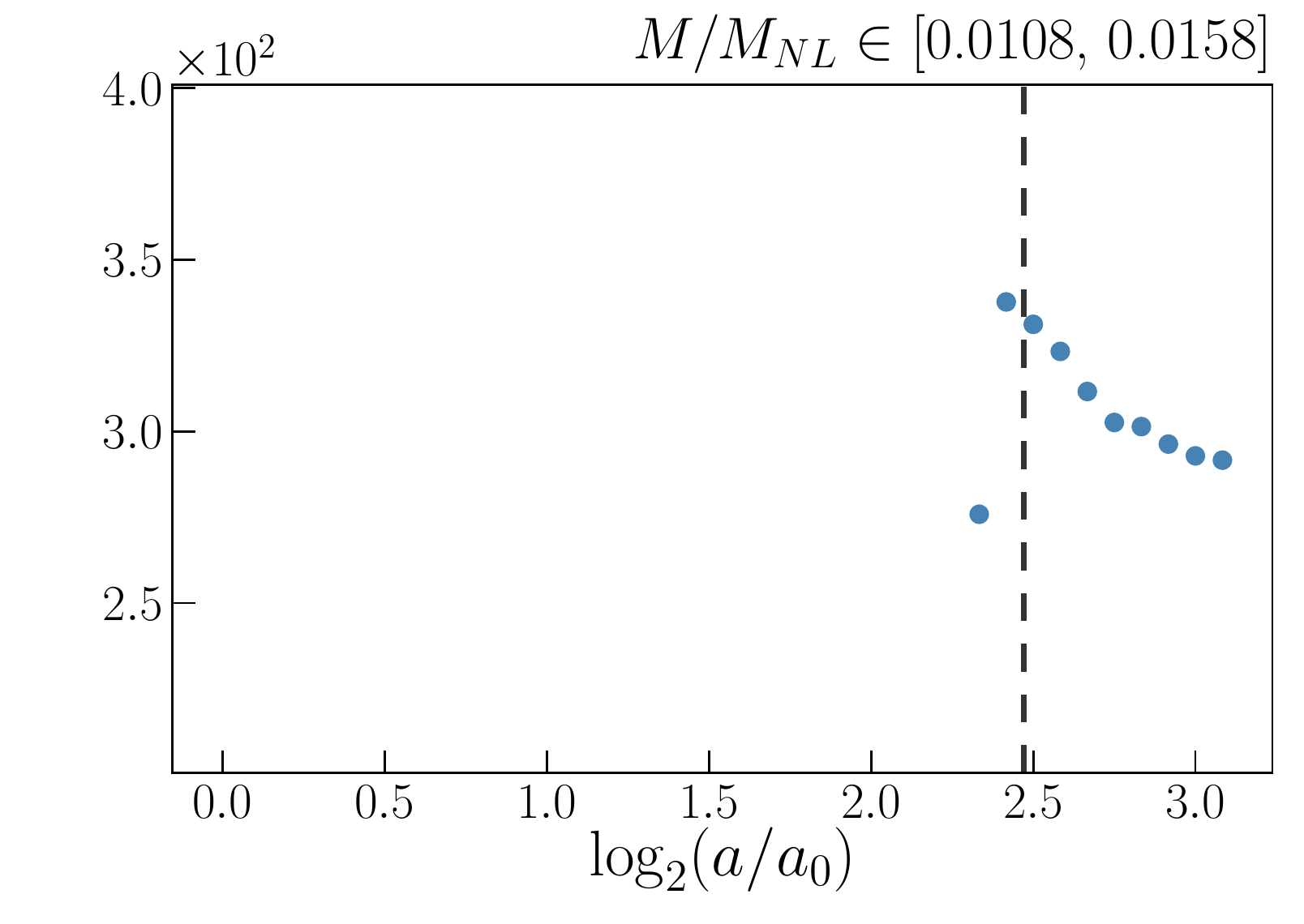}
    \end{subfigure}
    \hspace*{-0.7cm}
    \begin{subfigure}[b]{0.35\textwidth}
    \includegraphics[width=0.9\textwidth,height=0.6375\linewidth]{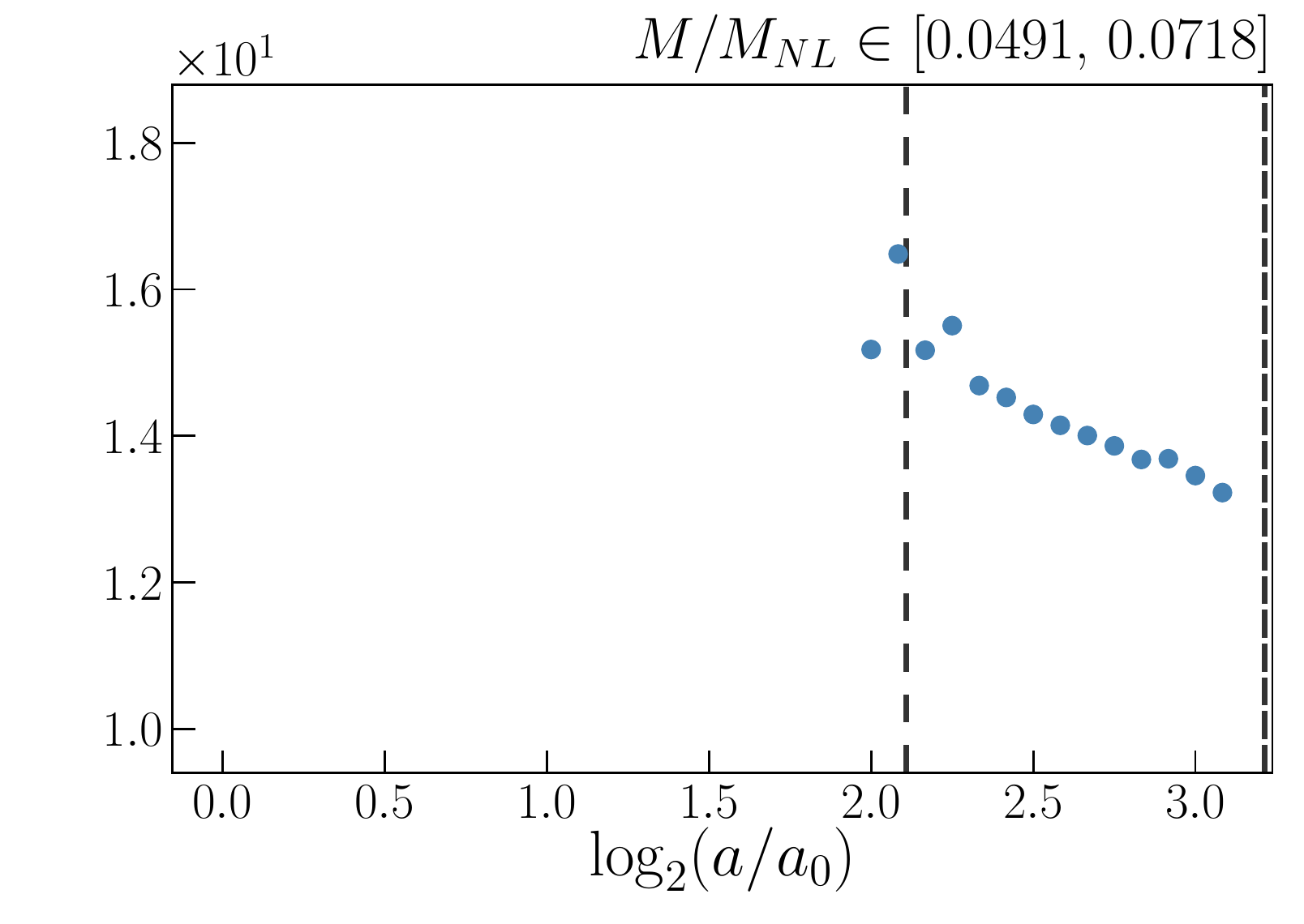}
    \end{subfigure}
    \\
    \centering
    \begin{subfigure}[b]{0.35\textwidth}
    \includegraphics[width=0.9\textwidth,height=0.6375\linewidth]{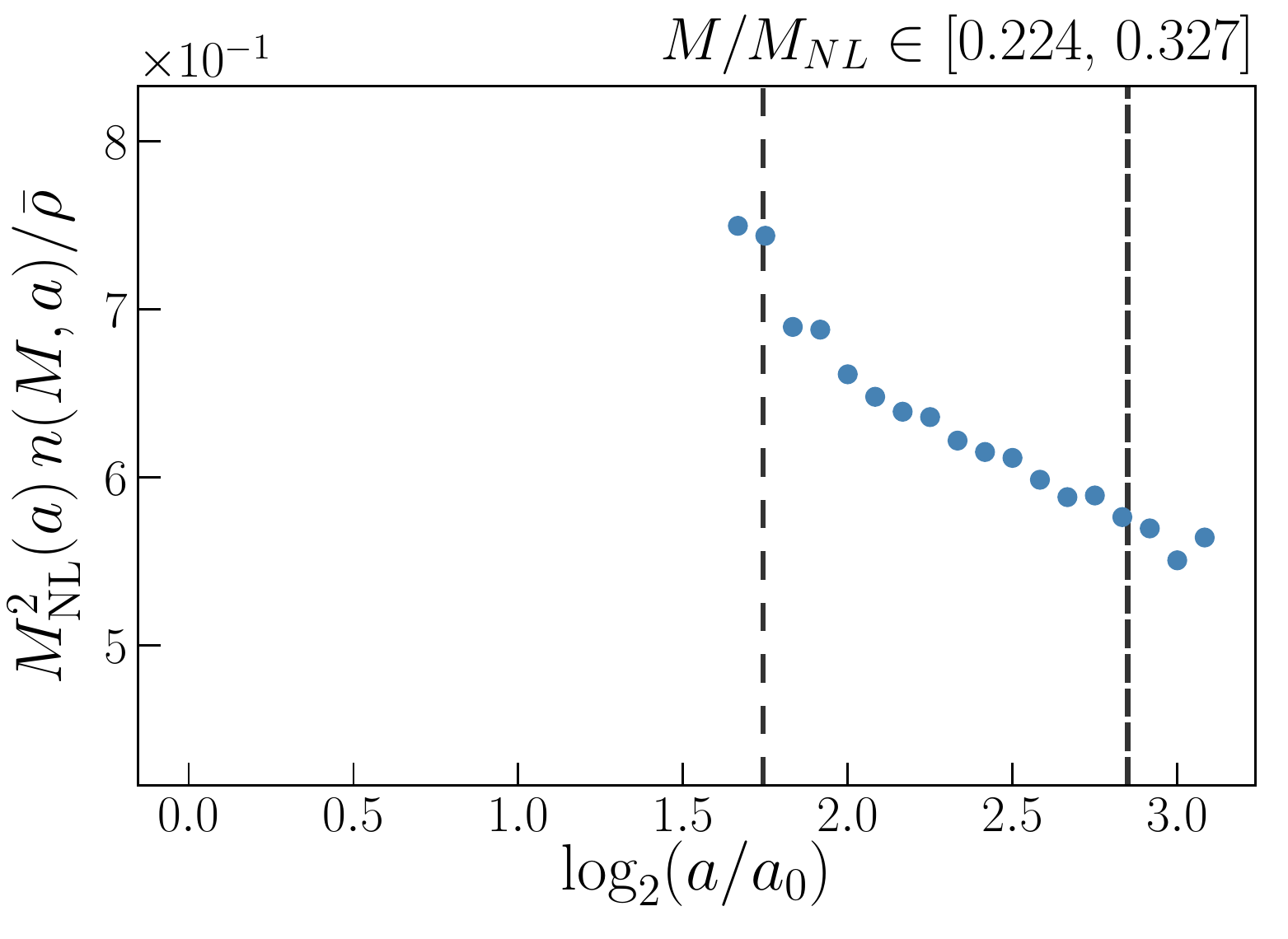}
    \end{subfigure}
    \hspace*{-0.7cm}
    \begin{subfigure}[b]{0.35\textwidth}
    \includegraphics[width=0.9\textwidth,height=0.6375\linewidth]{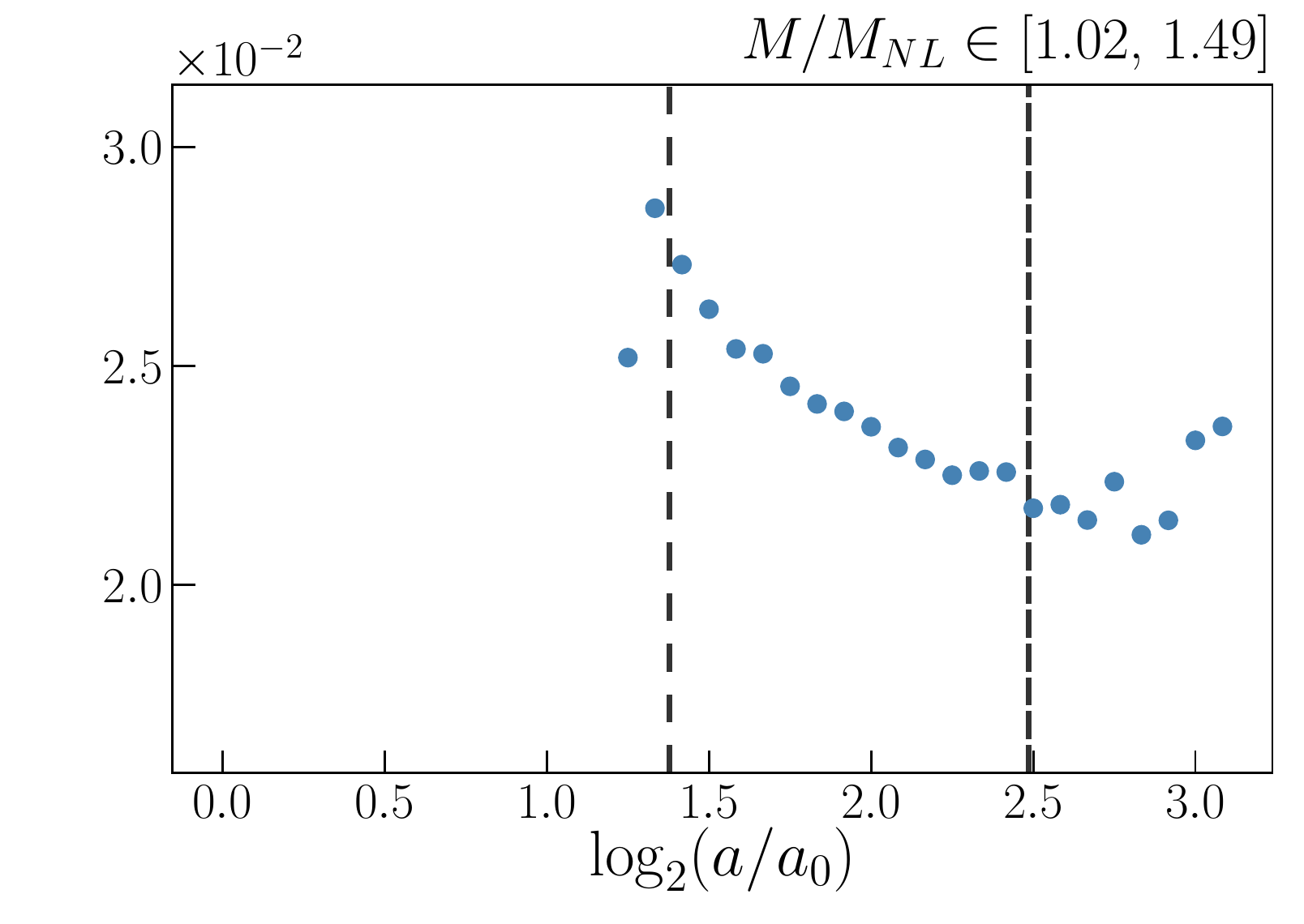}
    \end{subfigure}
    \hspace*{-0.7cm}
    \begin{subfigure}[b]{0.35\textwidth}
    \includegraphics[width=0.9\textwidth,height=0.6375\linewidth]{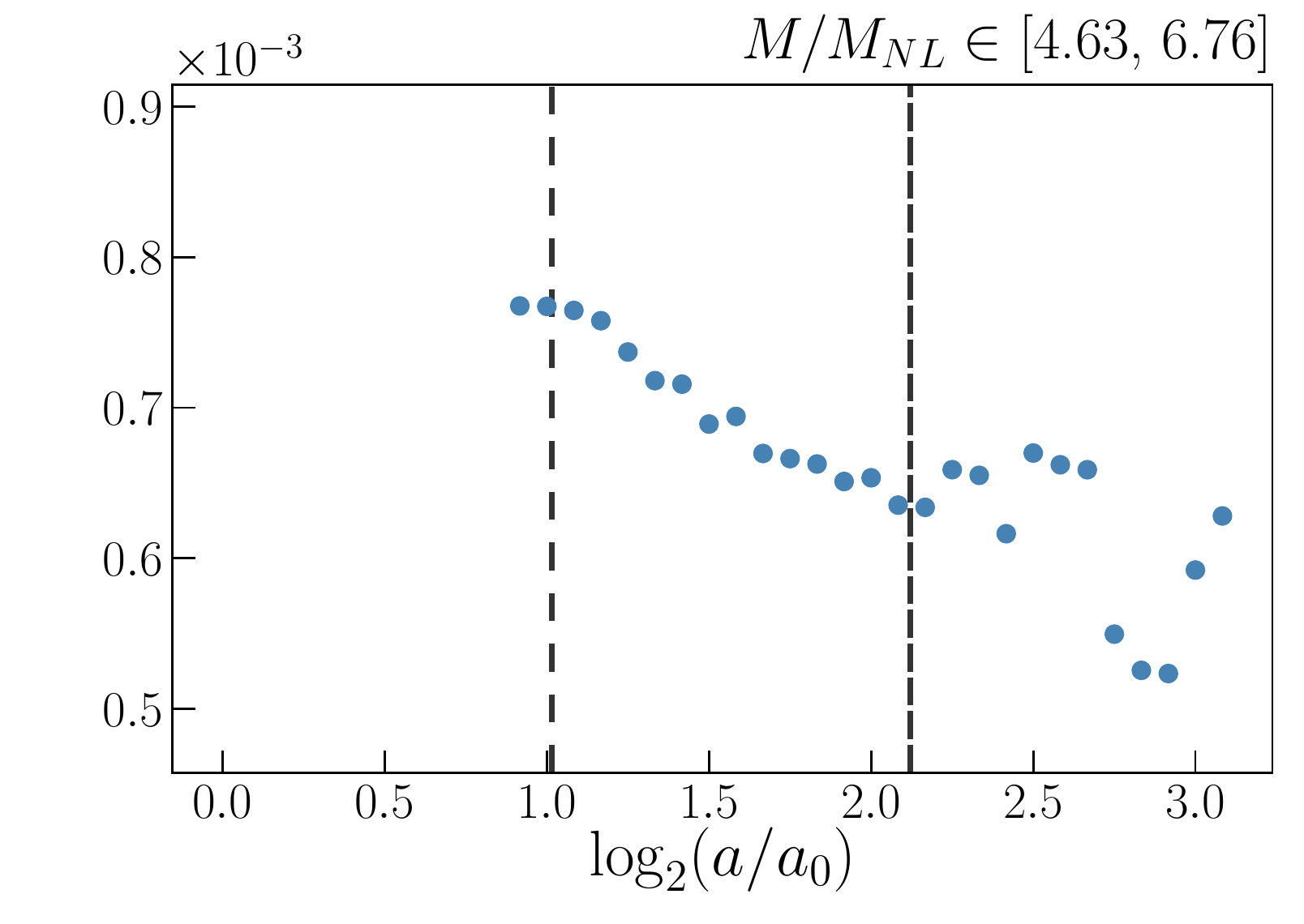}
    \end{subfigure}
    \\
    \centering
    \begin{subfigure}[b]{0.35\textwidth}
    \includegraphics[width=0.9\textwidth,height=0.6375\linewidth]{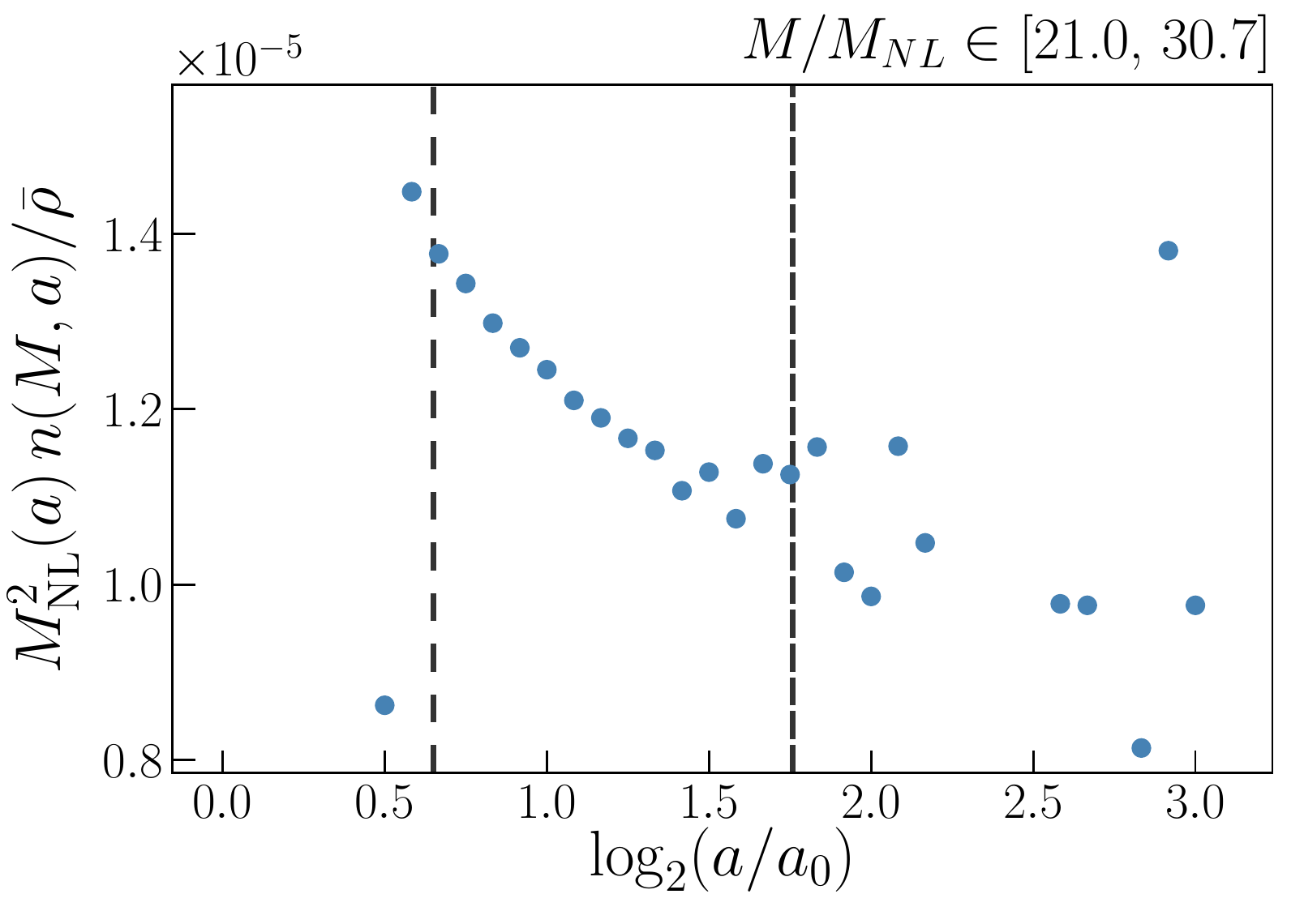}
    \end{subfigure}
    \hspace*{-0.7cm}
    \begin{subfigure}[b]{0.35\textwidth}
    \includegraphics[width=0.9\textwidth,height=0.6375\linewidth]{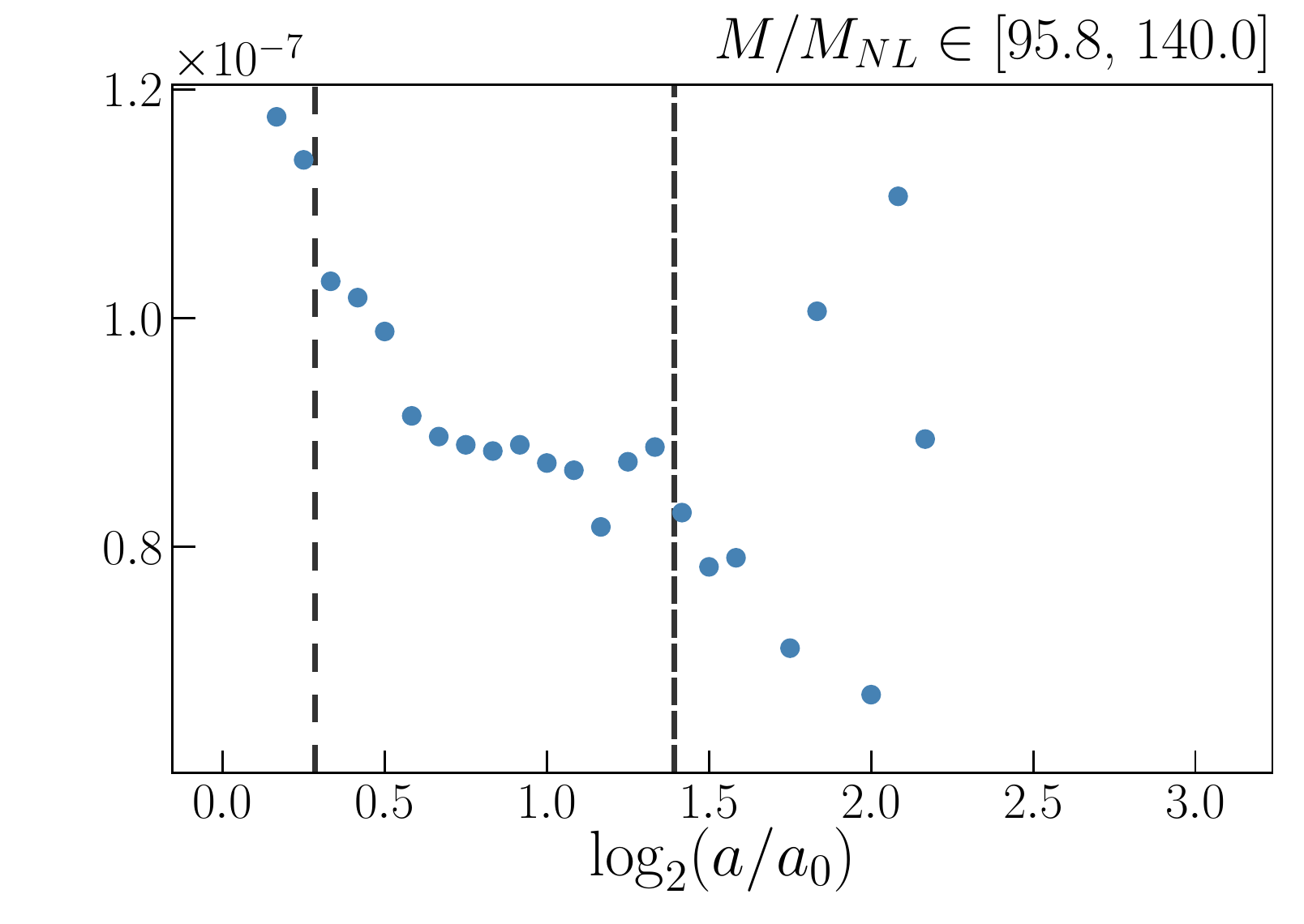}
    \end{subfigure}
    \hspace*{-0.7cm}
    \begin{subfigure}[b]{0.35\textwidth}
    \includegraphics[width=0.9\textwidth,height=0.6375\linewidth]{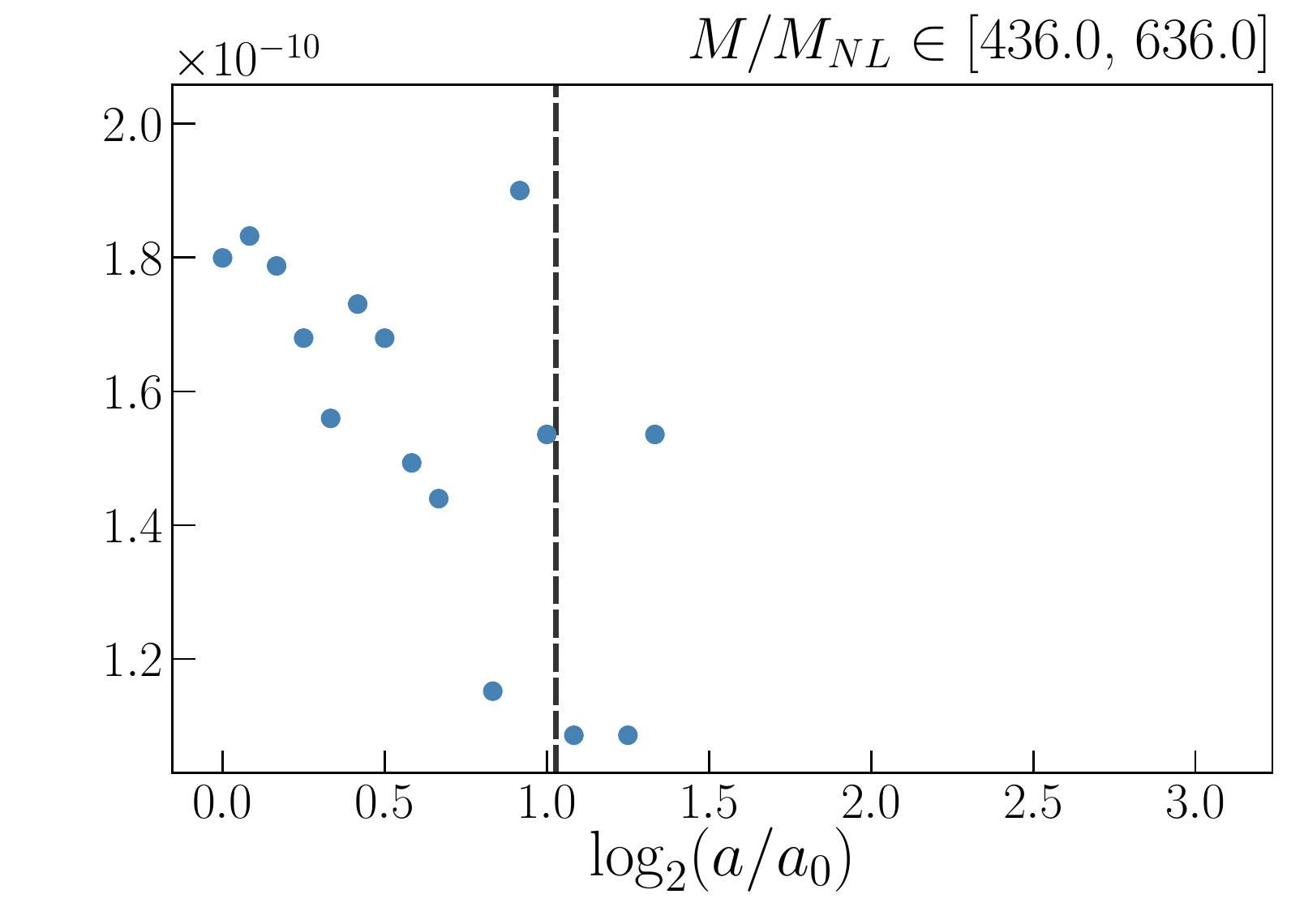}
    \end{subfigure}
\caption{Dependence on resolution of the \FOF halo mass function. Each plot corresponds to the bin of rescaled mass $M/M_{\mathrm{NL}}$ indicated on the top right and shows the rescaled halo mass function as a function of scale-factor. The loosely (densely) dashed black vertical line indicates the time at which the central mass in the bin corresponds to 50 (5000) particles. 
The snapshot spacing is such that this number increases by a factor 
of two for every two snapshots. We observe at best marginal evidence for convergence to a resolution independent value in a few of the bins, starting from
of order 5000 particles per halo. The behaviour can be contrasted with that in the following figure for \Rockstar halos. Note that, to facilitate comparison, 
the logarithmic range plotted on the $y$-axis is the same in all panels
and in both figures ($y_{\rm max}/y_{\rm min}=2$).}
\label{fig:FOF_HMF_CONVERGENCE}
\end{figure*}

\begin{figure*}
    \centering
    \begin{subfigure}[b]{0.35\textwidth}
    \includegraphics[width=0.9\textwidth,height=0.85\linewidth]{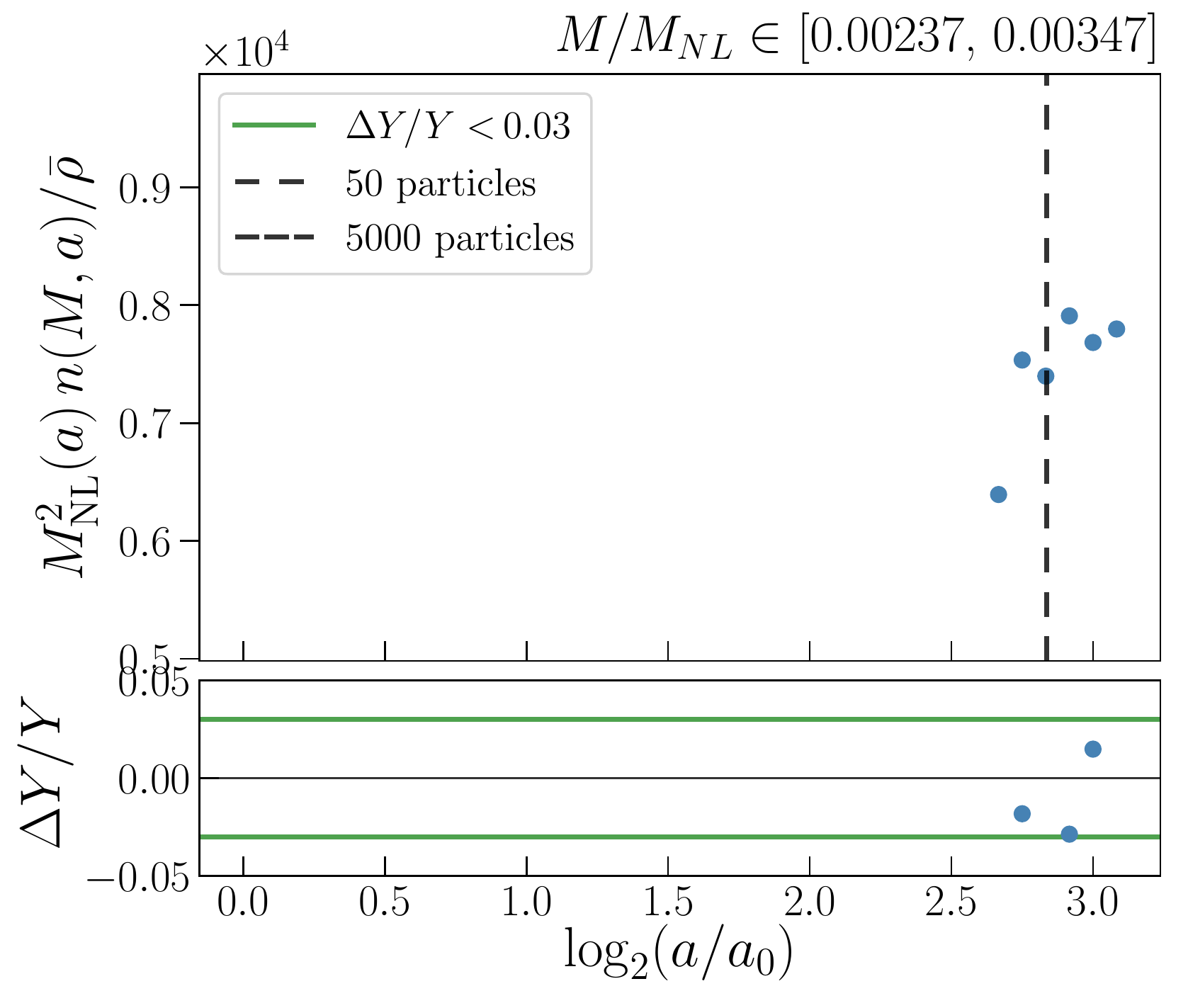}
    \end{subfigure}
    \hspace*{-0.7cm}
    \begin{subfigure}[b]{0.35\textwidth}
    \includegraphics[width=0.9\textwidth,height=0.85\linewidth]{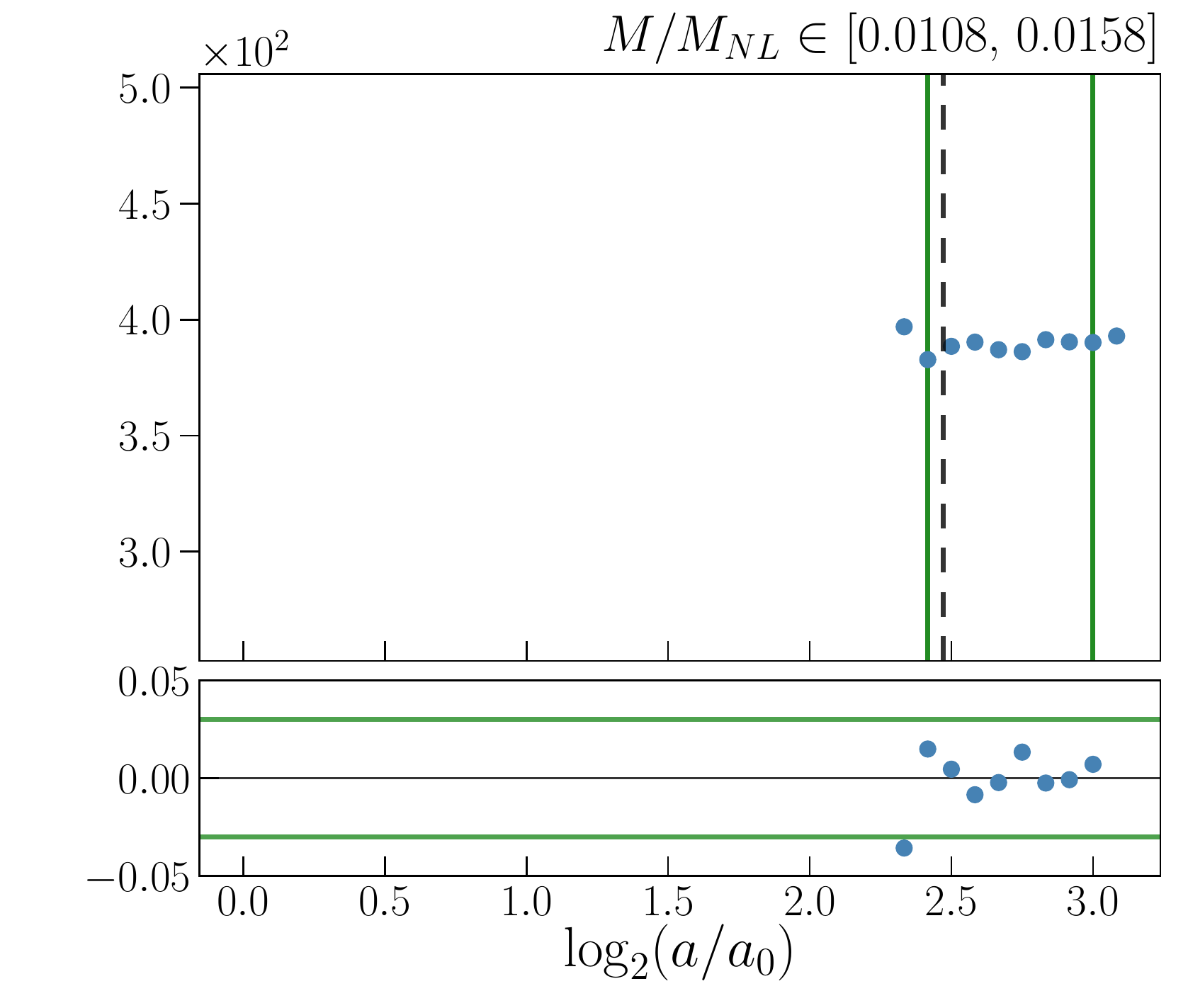}
    \end{subfigure}
    \hspace*{-0.7cm}
    \begin{subfigure}[b]{0.35\textwidth}
    \includegraphics[width=0.9\textwidth,height=0.85\linewidth]{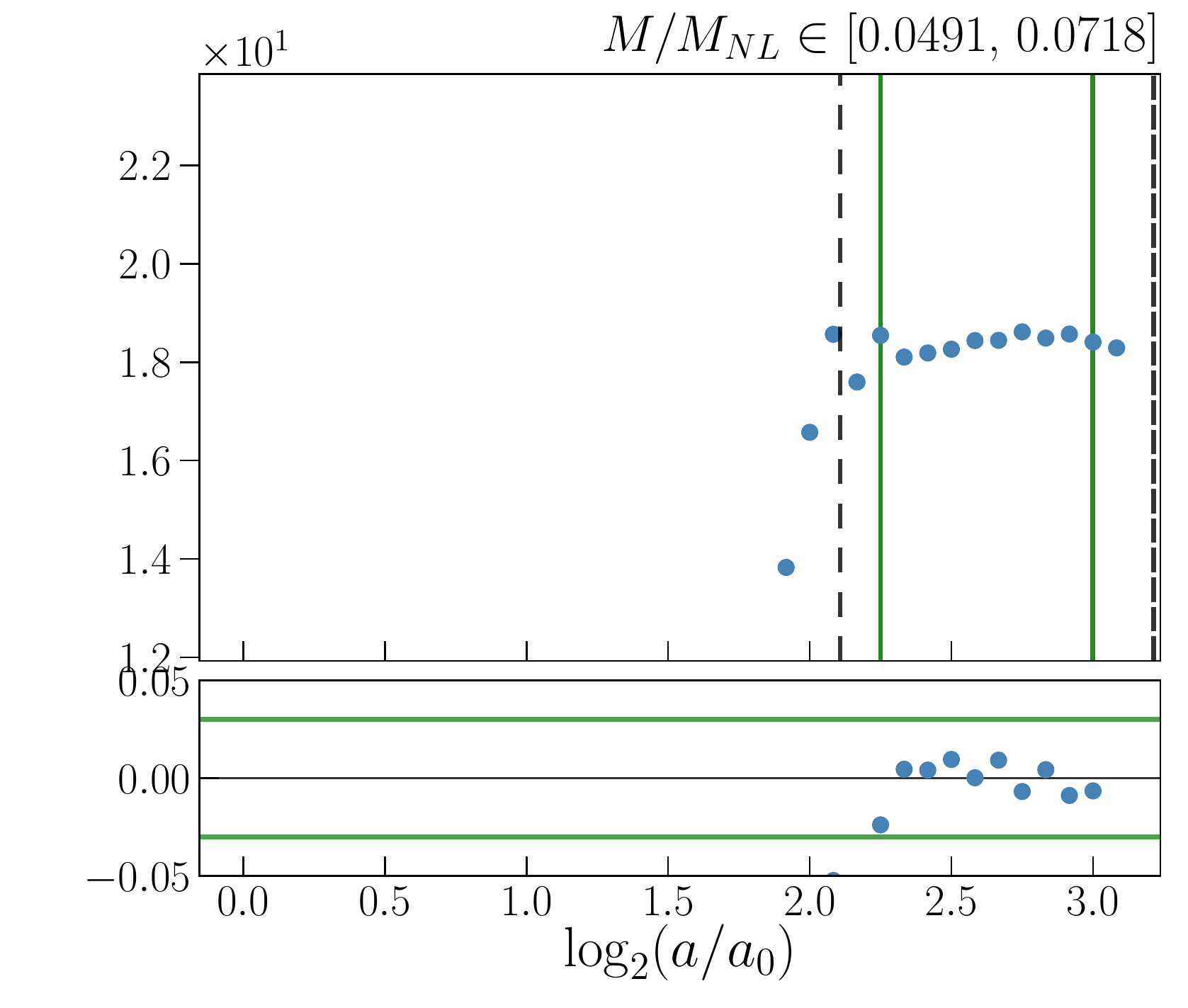}
    \end{subfigure}    \\
    \centering
    \begin{subfigure}[b]{0.35\textwidth}
    \includegraphics[width=0.9\textwidth,height=0.85\linewidth]{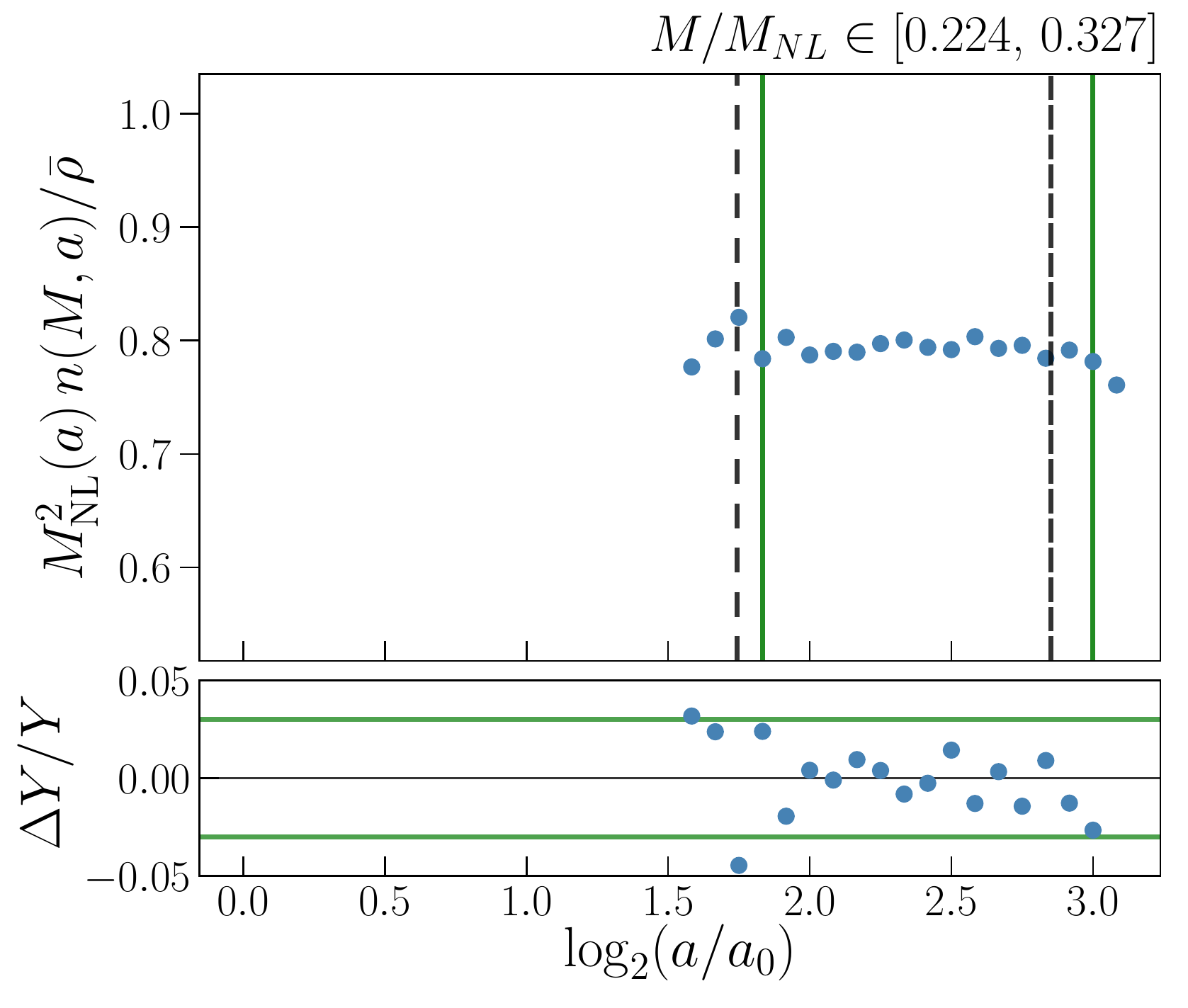}
    \end{subfigure}
    \hspace*{-0.7cm}
    \begin{subfigure}[b]{0.35\textwidth}
    \includegraphics[width=0.9\textwidth,height=0.85\linewidth]{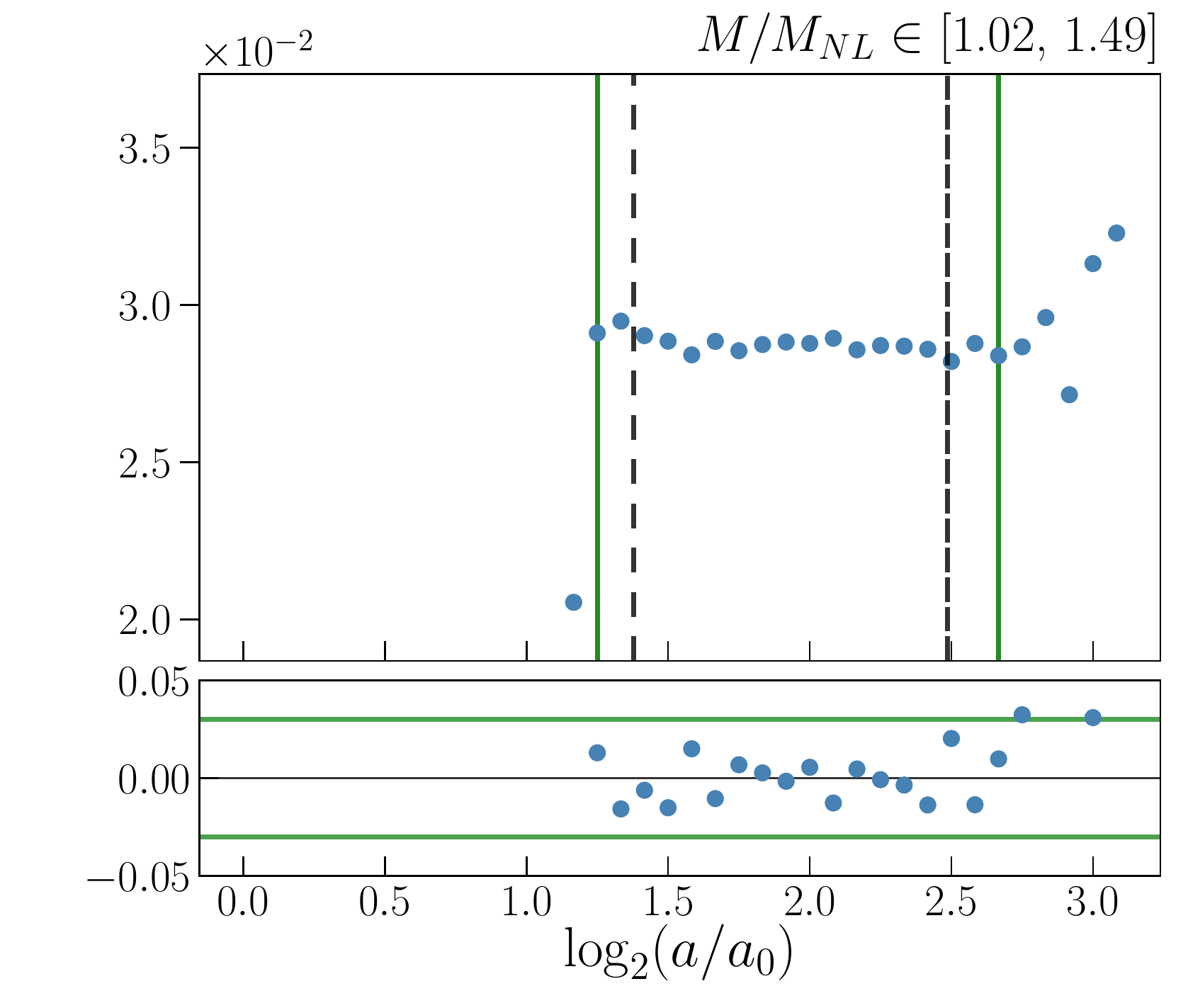}
    \end{subfigure}
    \hspace*{-0.7cm}
    \begin{subfigure}[b]{0.35\textwidth}
    \includegraphics[width=0.9\textwidth,height=0.85\linewidth]{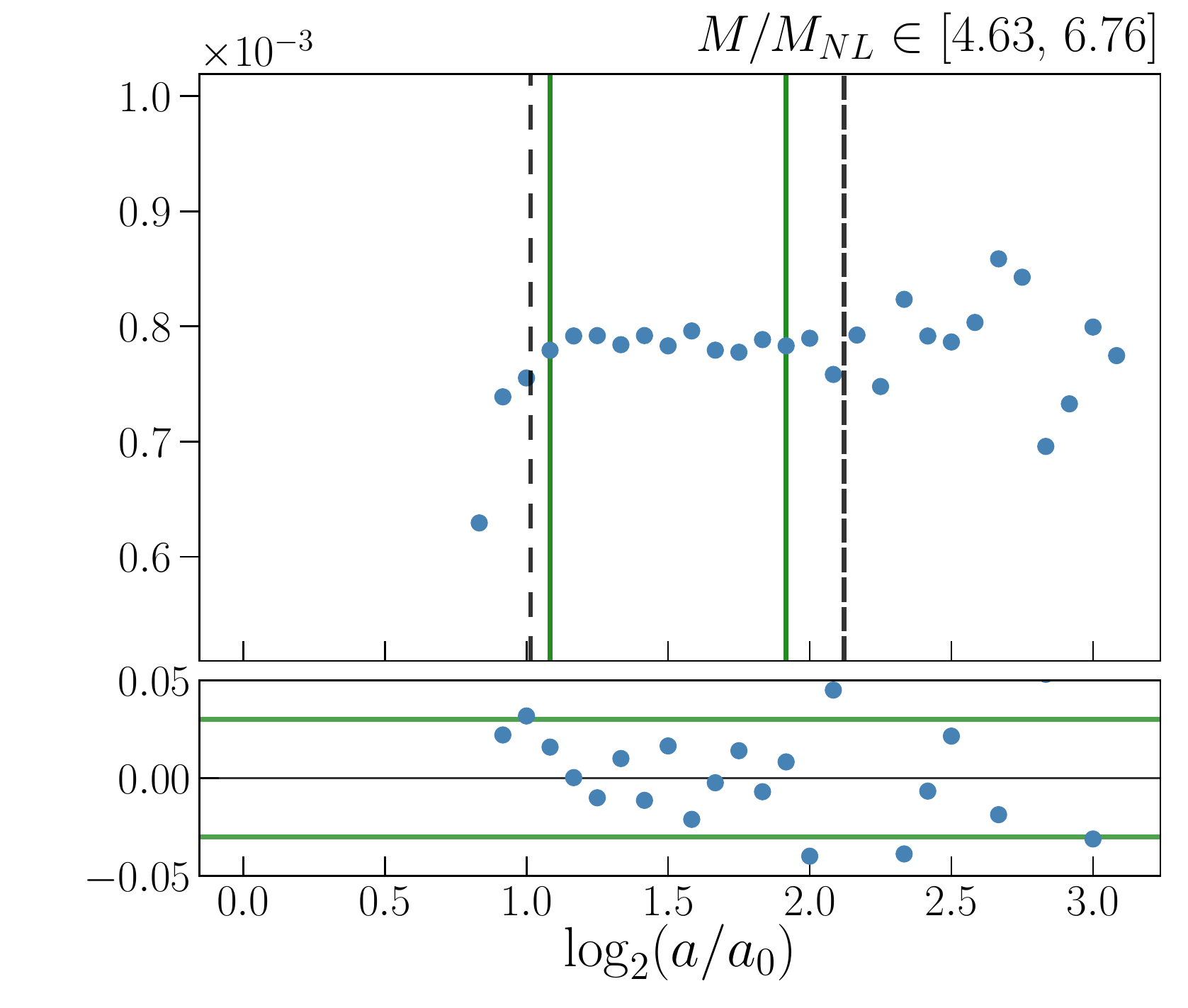}
    \end{subfigure}
    \\
    \centering
    \begin{subfigure}[b]{0.35\textwidth}
    \includegraphics[width=0.9\textwidth,height=0.85\linewidth]{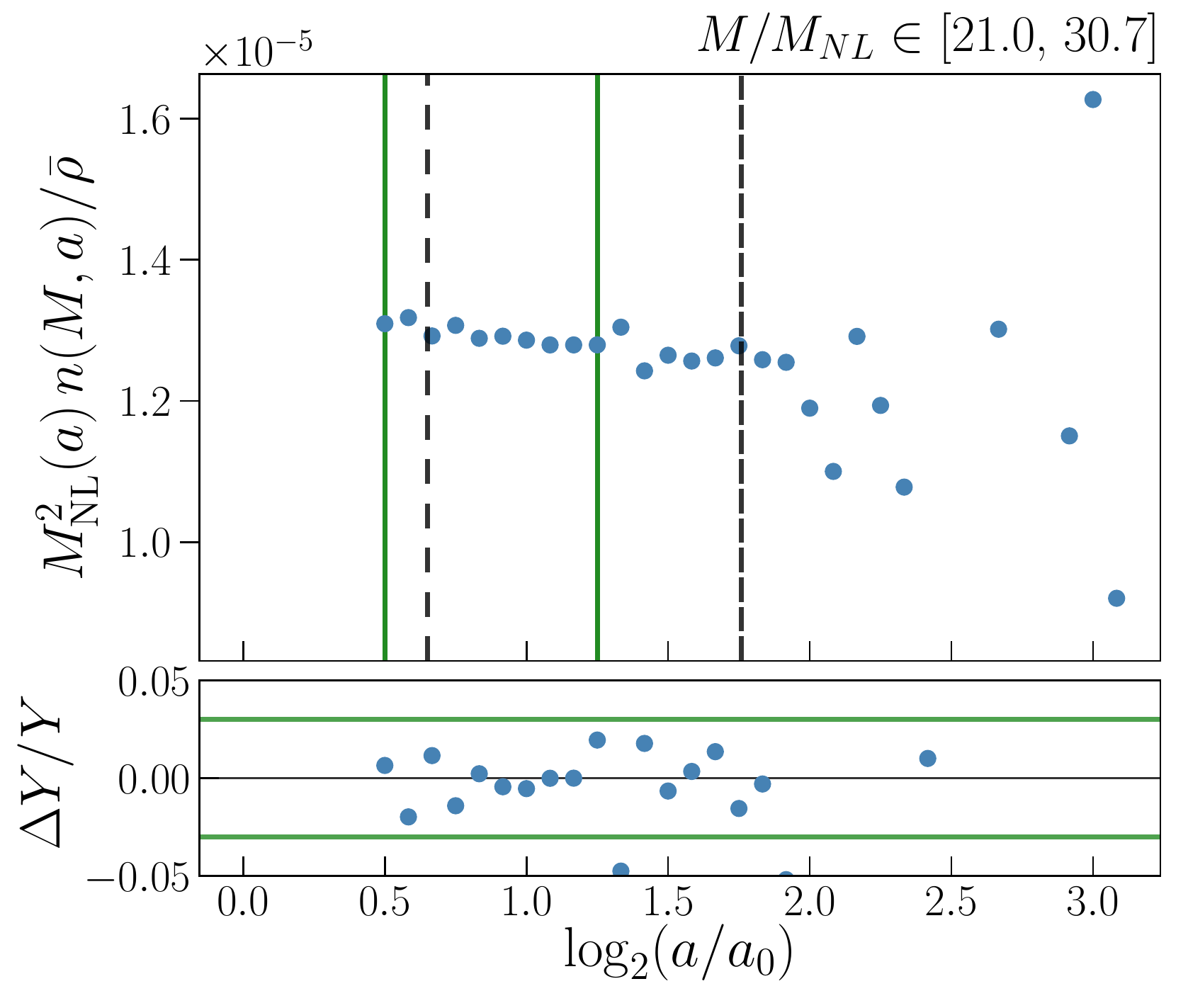}
    \end{subfigure}
    \hspace*{-0.7cm}
    \begin{subfigure}[b]{0.35\textwidth}
    \includegraphics[width=0.9\textwidth,height=0.85\linewidth]{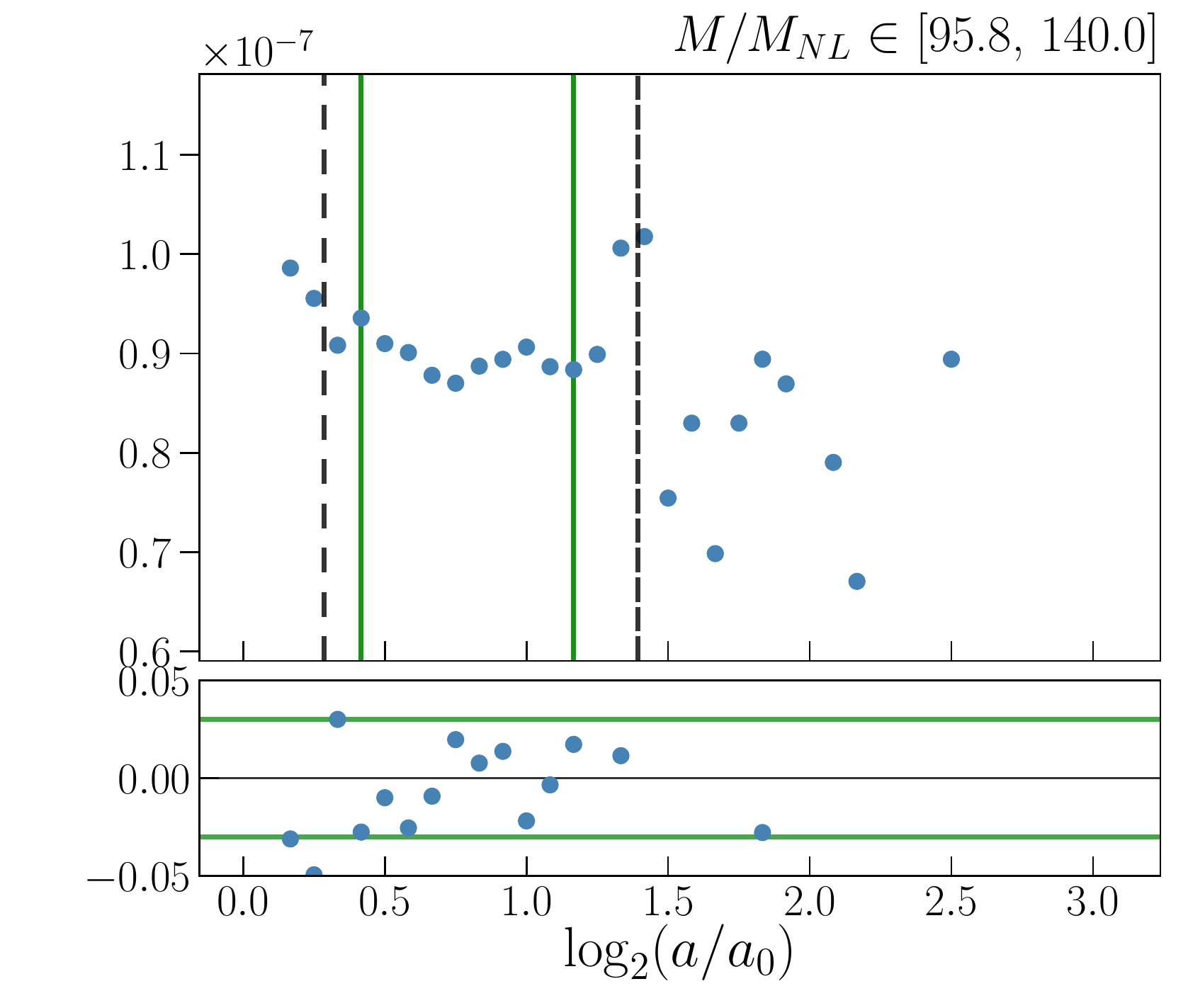}
    \end{subfigure}
    \hspace*{-0.7cm}
    \begin{subfigure}[b]{0.35\textwidth}
    \includegraphics[width=0.9\textwidth,height=0.85\linewidth]{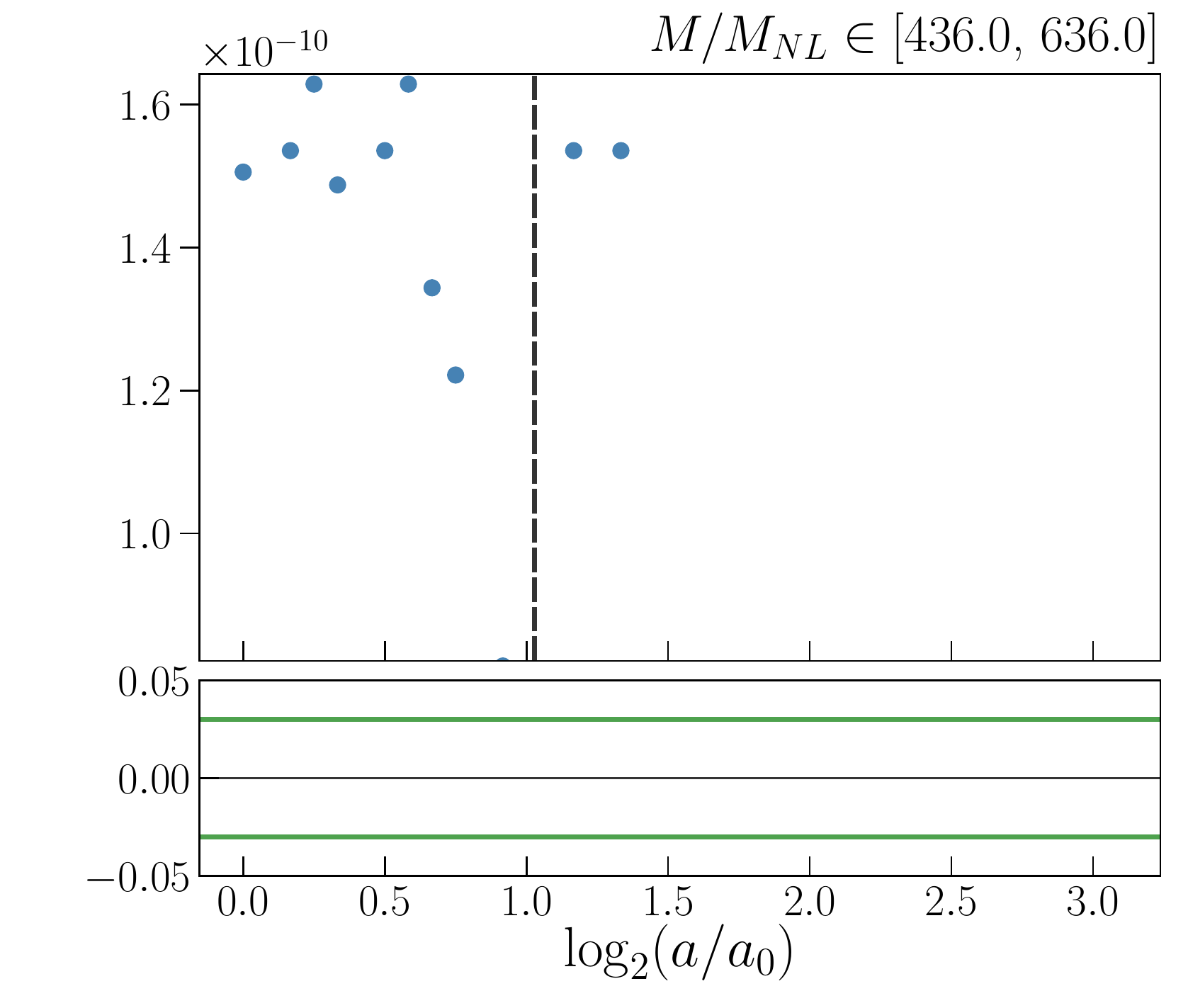}
    \end{subfigure}
\caption{Dependence on resolution of the \Rockstar halo mass function. As in the previous figure, except that a sub-panel has been added in every plot 
the fractional change $\Delta Y/Y$ of the rescaled mass function relative to the next snapshot. The full vertical green lines in the upper panels correspond to 
the most extended regions in which $\Delta Y/Y$ is below the indicated 
values, also shown by vertical green lines in the lower panel. In contrast to the previous figure (in which the logarithmic range on the $y$-axis is identical) we observe very clear evidence for convergence to a resolution independent result in almost all bins starting from around 50 particles per halo. The green vertical lines indicate the converged regions defined as the largest contiguous range of snapshots in which $\Delta Y/Y$ (i) lies in the range indicated by the horizontal green lines, and (ii) changes sign at least once.}
\label{fig:ROCKSTAR_HMF_CONVERGENCE}
\end{figure*} 

These figures show that the self-similar scaling appears to apply to a very good approximation to both \FOF and \Rockstar catalogs., but suggest that this scaling may be more closely followed in the latter catalog. To assess the degree of this self-similarity quantitatively, we follow the method defined in P1 and consider the rescaled HMF at fixed values of $M/M_{\mathrm{NL}}$ as a function of $\log_{2}(a/a_{0})$.
This corresponds to a projection of Figures \ref{fig:FOF_HMF} and \ref{fig:ROCKSTAR_HMF} along the x-axis. The results are shown in Figure \ref{fig:FOF_HMF_CONVERGENCE} for the
\FOF halos, and in Figure \ref{fig:ROCKSTAR_HMF_CONVERGENCE} for the \Rockstar halos. 
Each panel shows the measured value of the rescaled mass function for all 38 snapshots for a finite bin of halo mass $M/M_{\mathrm{NL}}$.
The bins have an equal logarithmic spacing with a width of approximately $40\%$ their central value, and the subsample of bins shown in the figures are equally spaced across the full range of $M/M_{\mathrm{NL}}$ sampled by the simulation. 
To facilitate comparison, the $y$-axis in all panels of the 
two figures has been chosen to have the same logarithmic range, with
$y_{\rm max}/y_{\rm min}=2$.
For the \Rockstar halos each plot also has a small subplot showing the fractional change $\Delta Y/Y$ of $Y=M_{\mathrm{NL}}^2 n(M,a)$ between consecutive snapshots i.e. at each $a_i$ we plot
$\Delta Y/Y=(Y_i/Y_{i-1})-1$.

Given that the range of $M/M_{\mathrm NL}$ is constant in each plot, 
and that $M_{\mathrm NL}$ expressed in units of particle mass increases 
monotonically as a function of time, the $x$-axis of these
plots can be labelled also by the monotonically increasing 
number of particles per halo. Thus each plot effectively shows
the measured mass function as a function of increasing resolution. 
The dashed vertical lines in each plot 
indicate a resolution corresponding to 50 and 5000 
particles respectively (for the geometric centre of 
the bin of $M_{\mathrm NL}$). At any intermediate time
the number of particles can easily be read off given 
our chosen snapshot spacing Eq.~(\ref{eq-snapshot-spacing}).
There are thus no halos in each plot until the time 
at which the particle number per halo reaches the minimal 
number of particles in halos included in the catalog (25 for
\FOF, 2 for \Rockstar), and it is only in the last bin, 
corresponding to the largest mass at given $M_{\mathrm NL}$, 
that there are halos at the first output time 
(i.e. at $a=a_0$ defined by Eq.~(\ref{a0-ref})).  

While in a given plot the number of particles per halo increases monotonically
from left to right, the average number of halos contributing to the measure of the HMF in the corresponding bin decreases monotonically: this number is proportional to the simulation volume in units of the characteristic volume $R_{\mathrm NL}^3$,
so, in the approximation that $M_{\mathrm{NL}}^2 n(M,a)$  is constant,
it is proportional to $1/M_{\mathrm NL}$. Thus we expect the 
effect of sparseness of sampling in the finite bins to manifest itself
as increasing noise in the measured signal at later times. This is indeed  
what we see in Figures \ref{fig:FOF_HMF_CONVERGENCE} and \ref{fig:ROCKSTAR_HMF_CONVERGENCE}, for the larger values of 
$M/M_{\mathrm{NL}}$ (for which the number of halos per
logarithmic interval of mass decreases strongly). Conversely 
we see that most of our plots are clearly not dominated 
by such sampling noise, despite our use of rather narrow mass bins.
We can thus clearly identify systematic dependencies on 
resolution alone.

\subsection{FOF halos}

We see in Figure \ref{fig:FOF_HMF_CONVERGENCE} that once the time is reached at which the mass range in the bin is unaffected by the intrinsic lower limit on the number of particles in the catalog, the rescaled \FOF HMF apparently decreases monotonically as the number of particles per halo increases, until it becomes noisy due to sparseness effects at the later times. 
There is marginal evidence, at best, for convergence towards a resolution independent value in a few of the mass bins. In these cases the apparent 
flattening out of the curves occurs at several thousand particles. For halos of $50-100$ particles, the inferred converged value is then systematically overestimated by $20-25\%$.
  
This observed monotonic decrease in the region where the mass function is well measured has an obvious interpretation in terms of resolution dependence of the mass measured by the \FOF algorithm. Indeed studies in the literature \citep{warren_et_al_2006,2011ApJS..195....4M} have shown, by using the \FOF algorithm on idealized isolated halos, that it systematically overestimates mass because of finite size effects that lead to percolation into sub-critical density regions. As the underlying mass function is a decreasing function of mass, the number of halos for a given measured \FOF mass are 
as a result overestimated at lower resolution, as observed in Figure \ref{fig:FOF_HMF_CONVERGENCE}. Further the order of magnitude of the 
overestimate we observe is consistent with the effects quantified on isolated halos by \cite{warren_et_al_2006} and \cite{2011ApJS..195....4M}.
Our method of identifying and quantifying these resolution effects, using deviations from self-similarity,
has the considerable advantage of allowing its precise ``in situ" quantification in a cosmological simulation.

\subsection{Rockstar halos}

For \Rockstar halos, Figure \ref{fig:ROCKSTAR_HMF_CONVERGENCE} shows that the rescaled mass function is much more self-similar than for \FOF. In all 
but the first bin, the data clearly shows convergence in a range
of scale-factors: the visually apparent plateau in each case corresponds,
as can be seen in the lower panel, to fluctuations of the derivative 
which appear to be roughly symmetric about zero. Down to the intrinsic
limit on precision corresponding to these fluctuations (of order $2\%$ here), 
we thus conclude that there is convergence to a resolution 
independent physical value of the mass function.

We observe further that the point of onset of this convergence, right across the 
several decades of $M/M_{NL}$, corresponds to a number of particles per halo 
situated around $50$ particles, and at most $100$. For the larger mass 
bins we see both the growth in the fluctuations due to finite sampling
and also strong systematic deviations from the converged plateau. 
The latter are a direct manifestation of the finite box size: as shown in P1 using the analysis of the matter correlation function, by $\log_2(a/a_0) \approx 2.5$ the finite size of the simulation box begins to give rise to strong systematic deviations from self-similarity even at non-linear scales.

\begin{figure}
\centering\resizebox{8cm}{!}{\includegraphics[]{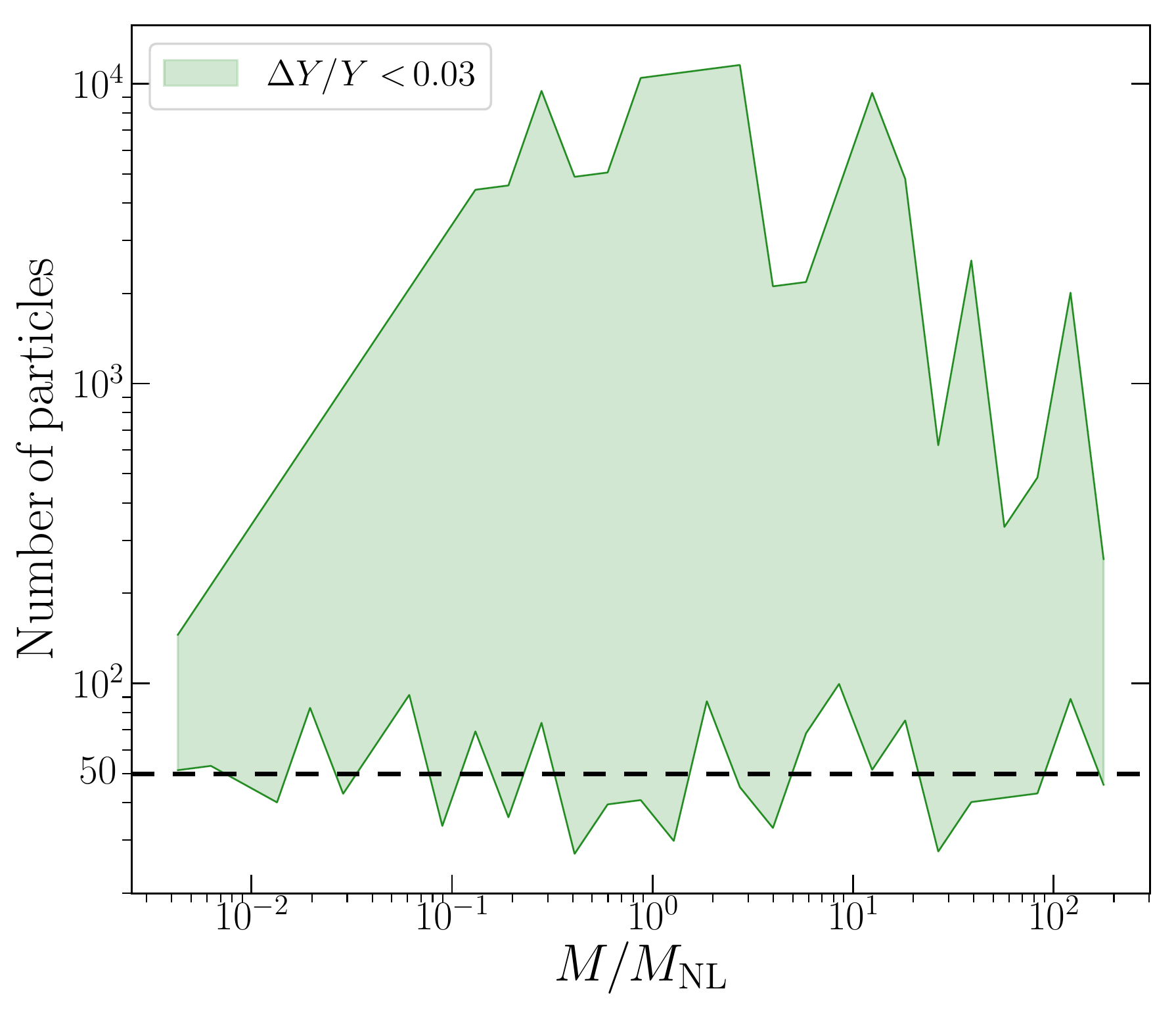}}
\caption{Range of particle number per halo in which the \Rockstar HMF is converged.
The convergent region, expressed as a range of particle number per halo, is plotted
as a function of the rescaled mass $M/M_{\mathrm{NL}}$. This plot is for the same binning as used in Figure \ref{fig:ROCKSTAR_HMF_CONVERGENCE} (but shows all 40 bins). The upper and lower bounds correspond to those indicated
by the green vertical lines in the upper panels in Figure \ref{fig:ROCKSTAR_HMF_CONVERGENCE}, inferred by finding the largest contiguous set of snapshots in which the variation lies below the threshold value shown by the 
green horizontal lines in the lower panels, and in which $\Delta Y$ changes sign at least once.}
\label{fig:ROCKSTAR_HMF_RESOLUTION}
\end{figure}

To quantify these observations a little more precisely we plot in Figure \ref{fig:ROCKSTAR_HMF_RESOLUTION}, for all 40 bins of $M/M_{\mathrm{NL}}$
from which those shown in Figure \ref{fig:ROCKSTAR_HMF_CONVERGENCE} are sampled, the converged range expressed in terms of the particle number per halo (evaluated at the geometric centre of the bin). The range is determined here as the largest contiguous set of snapshots in which $\Delta Y/Y <0.03$, corresponding to the green-dashed lines in the sub-panels of Figure \ref{fig:ROCKSTAR_HMF_CONVERGENCE}, and in which 
$\Delta Y$ changes sign at least once. 
As anticipated, we see that the lower cut-off on particle number per halo we determined fluctuates in a range around $50$ particles up to at most $100$ particles.
The upper cut-offs in the plot quantify limits on the accurate 
determination of the HMF arising from the finite size
of the simulation box. The increasing upper cut-off at smaller
$M/M_{NL}$ reflect the fact that in any such bin the largest mass which can 
be potentially resolved is  $(M/M_{NL}) \times M_{NL}(a_f)$ 
where $a_f$ is the scale-factor at the stopping time of the simulation. 
This maximal mass thus grows linearly with $M/M_{NL}$ until, 
at some value of $M/M_{NL}$,  it reaches a value at which the 
sparseness of halos in the finite volume  starts make the measure 
of the very HMF noisy. The maximal value is then fixed as a function of time and thus decreases approximately as a function of $M/M_{NL}$, as observed. 

The chosen threshold value of $\Delta Y/Y$, inferred from the level of residual fluctuations in the subplots in Figure \ref{fig:ROCKSTAR_HMF_CONVERGENCE},  
gives a measure of the precision with which the mass function, in the chosen binning, can be determined. Varying the size of our bins we observe that
this precision level changes, reaching a minimal level for most bins of 
under one percent when we divide the full range of $M/M_{NL}$
into only a couple bins. Irrespective of these binnings we find that
we do not find any evidence for systematic evolution of the mass function
with resolution when the halos contain more than of order 100 particles.
On the other hand this statement becomes weaker as the residual 
fluctuations decrease below the few percent level 
because the bins themselves become very broad. Conservatively 
we thus conclude that the HMF is converged to within $3-5 \%$ of
its real value for halos with of order $50$ to $100$ particles.

\subsection{Comparison with other \Rockstar catalogs}
\label{Comparison-catalogs}

\begin{figure*}
    \begin{subfigure}[b]{0.35\textwidth}
    \includegraphics[width=0.9\textwidth,height=0.85\linewidth]{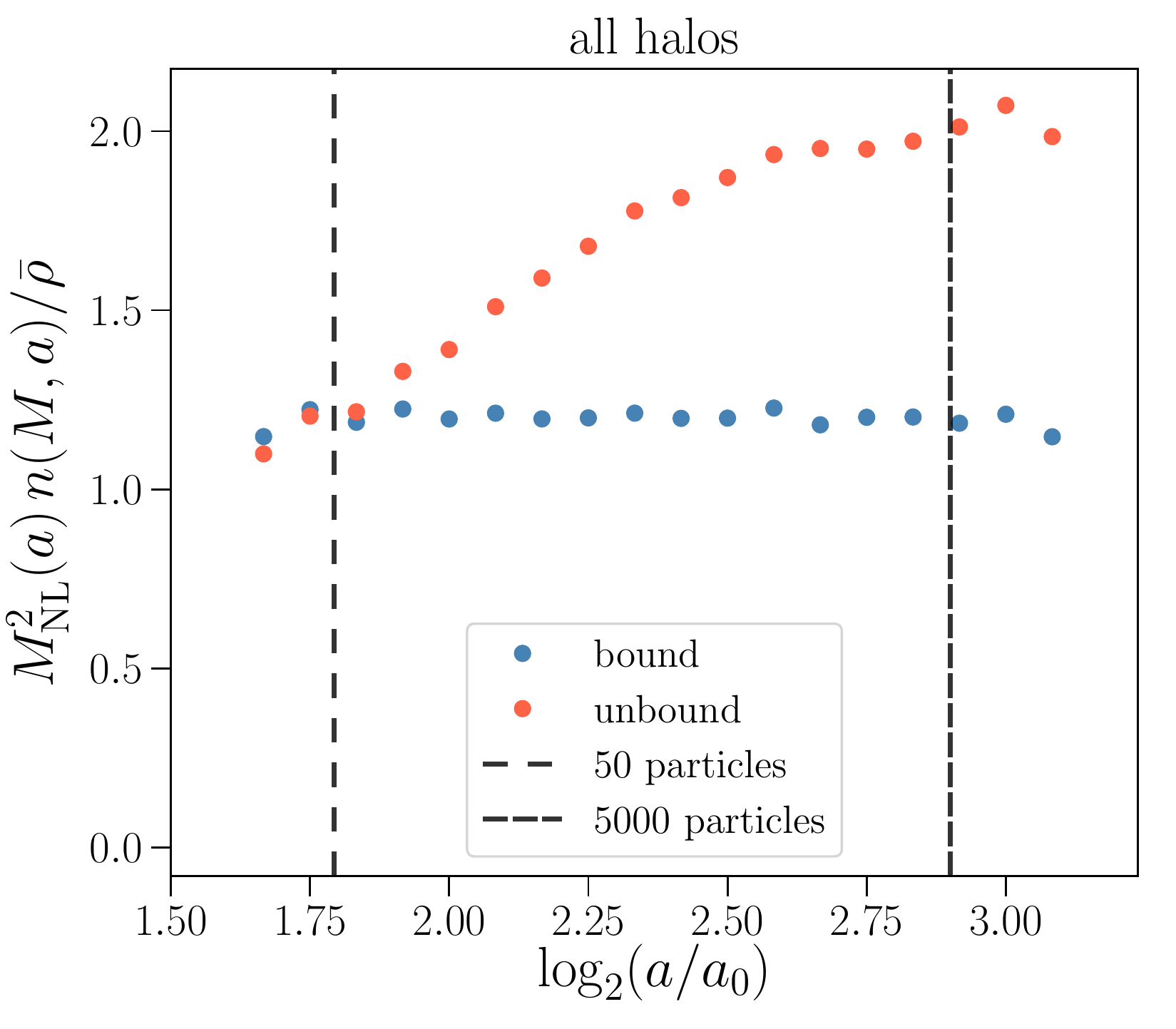}
    \end{subfigure}
    \hspace*{-0.7cm}
    \begin{subfigure}[b]{0.35\textwidth}
    \includegraphics[width=0.9\textwidth,height=0.85\linewidth]{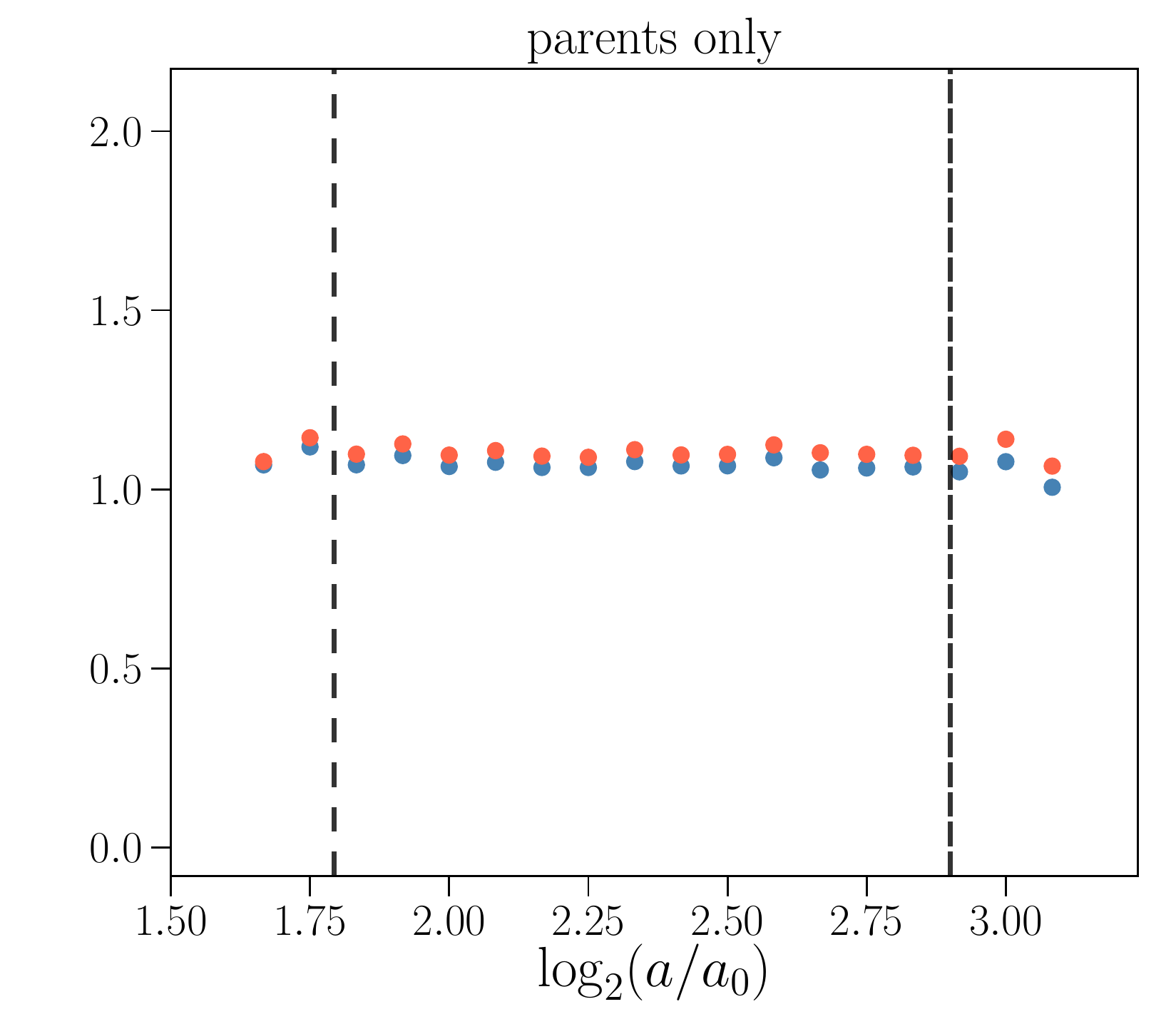}
    \end{subfigure}
    \hspace*{-0.7cm}
     \begin{subfigure}[b]{0.35\textwidth}
    \includegraphics[width=0.9\textwidth,height=0.85\linewidth]{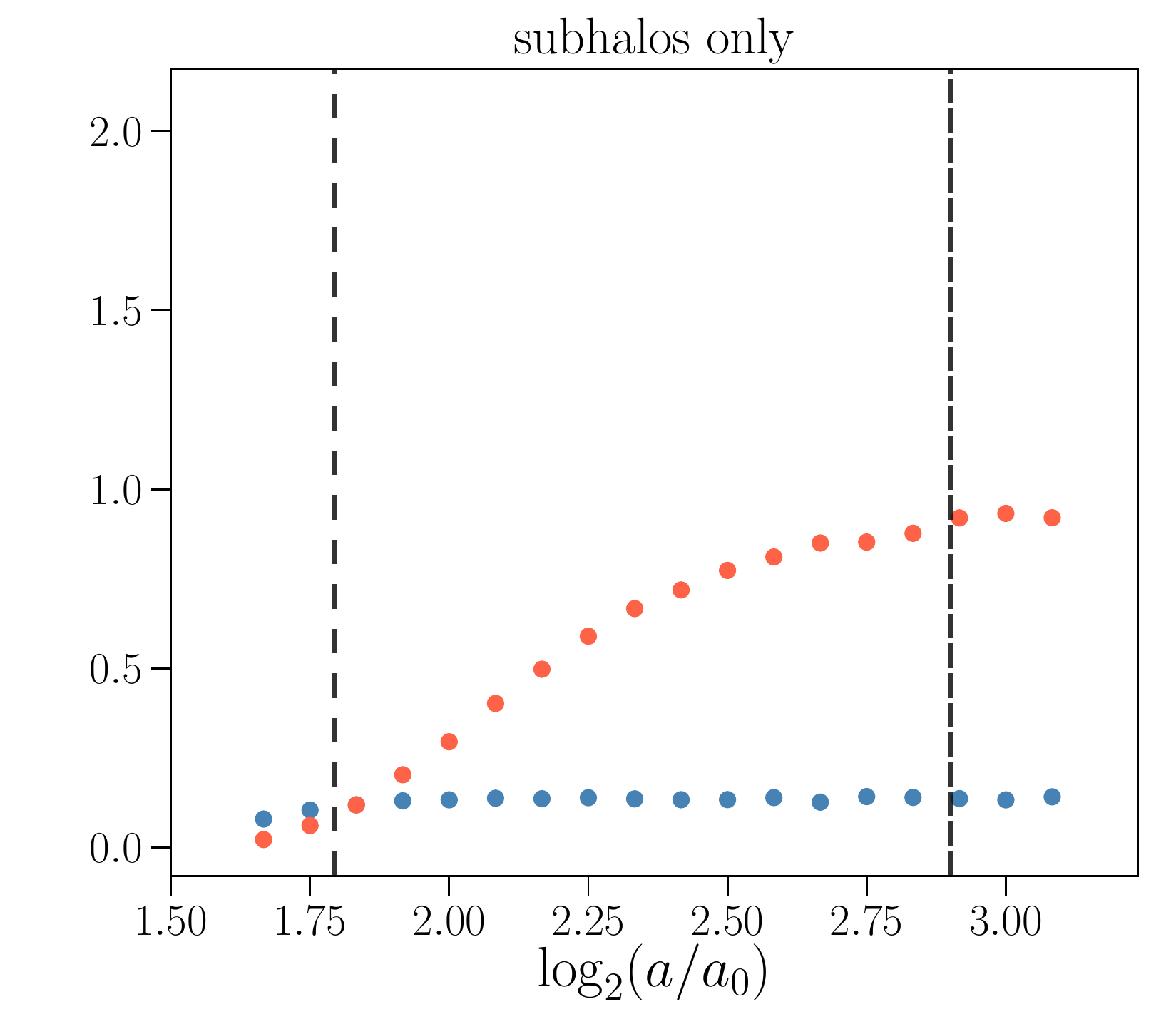}
    \end{subfigure}
\caption{Mass function convergence in different \Rockstar catalogs. Each panel shows a plot exactly like those in Figures \ref{fig:FOF_HMF_CONVERGENCE} and \ref{fig:ROCKSTAR_HMF_CONVERGENCE}, but
now for a single chosen rescaled mass bin with $M/M_{\mathrm{NL}} \in [0.20,0.25]$.
The left panel compares results for two catalogs containing all halos, one with the full masses obtained from \Rockstar's group finding and one with only the estimated gravitationally bound mass. The central panel compares these same two catalogs, but from which the subhalos have been removed. The right panel compares the subhalos with the two mass assignments.
The poor resolution of the underlying \FOF mass assignment is clearly 
corrected for by the removal of unbound mass. For this analysis all halos 
with less than 25 particles have been removed.}
\label{fig:ROCKSTAR_COMPARISON}
\end{figure*}

Our analysis clearly indicates that it is the resolution dependence of mass 
assignment in the \FOF algorithm that causes the poor accuracy in the determination of the HMF. As \Rockstar does so much better, the question obviously arises as to why there is such a great difference, given that \Rockstar is built itself on an initial \FOF selection. It is evident that a potentially important difference can arise from the mass unbinding performed by \Rockstar, which gives the final mass assigned to the halos
in the default output catalog we have analysed here.  As \Rockstar provides also output catalogs including both bound and unbound mass it is easy to test whether this is the case. 

Figure \ref{fig:ROCKSTAR_COMPARISON} shows a comparison between convergence plots, like those in Figures \ref{fig:FOF_HMF_CONVERGENCE} and \ref{fig:ROCKSTAR_HMF_CONVERGENCE}, but for a single chosen bin of rescaled
mass $M/M_{NL} \in [0.20,0.25]$. The left panel compares the catalog we 
have analysed above (bound mass only) with a catalog in which 
the same halos are assigned all their SO virial mass (bound and unbound). The central panel compares 
the same two catalogs but now including only the parent halos. The right panel 
shows the catalogs of subhalos only, with and without unbound mass. 

These plots show that removal of the unbound mass appears to be the essential step in the \Rockstar algorithm
which corrects for an otherwise strong resolution dependence in the mass function like that obtained in the  \FOF catalog. 
As can be seen from the second and third panels, this mass unbinding essentially affects only the population of 
subhalos while having negligible impact on the parent halos. We have compared in detail 
our quantitative results for convergence using the full catalog of bound SO masses  with those obtained using the 
parent only catalogs with the full SO mass and have found them to be essentially unchanged. Thus the good convergence 
could arguably be ascribed simply to the removal of the subhalos, rather than the mass unbinding. The third panel in Figure \ref{fig:ROCKSTAR_COMPARISON}, however,  
shows evidence for the recovery also of a converged mass function for subhalos due to mass unbinding.


\section{Correlation of halo centers}

\begin{figure*}
    \centering
    \begin{subfigure}[b]{0.35\textwidth}
    \includegraphics[width=0.9\textwidth,height=0.85\linewidth]{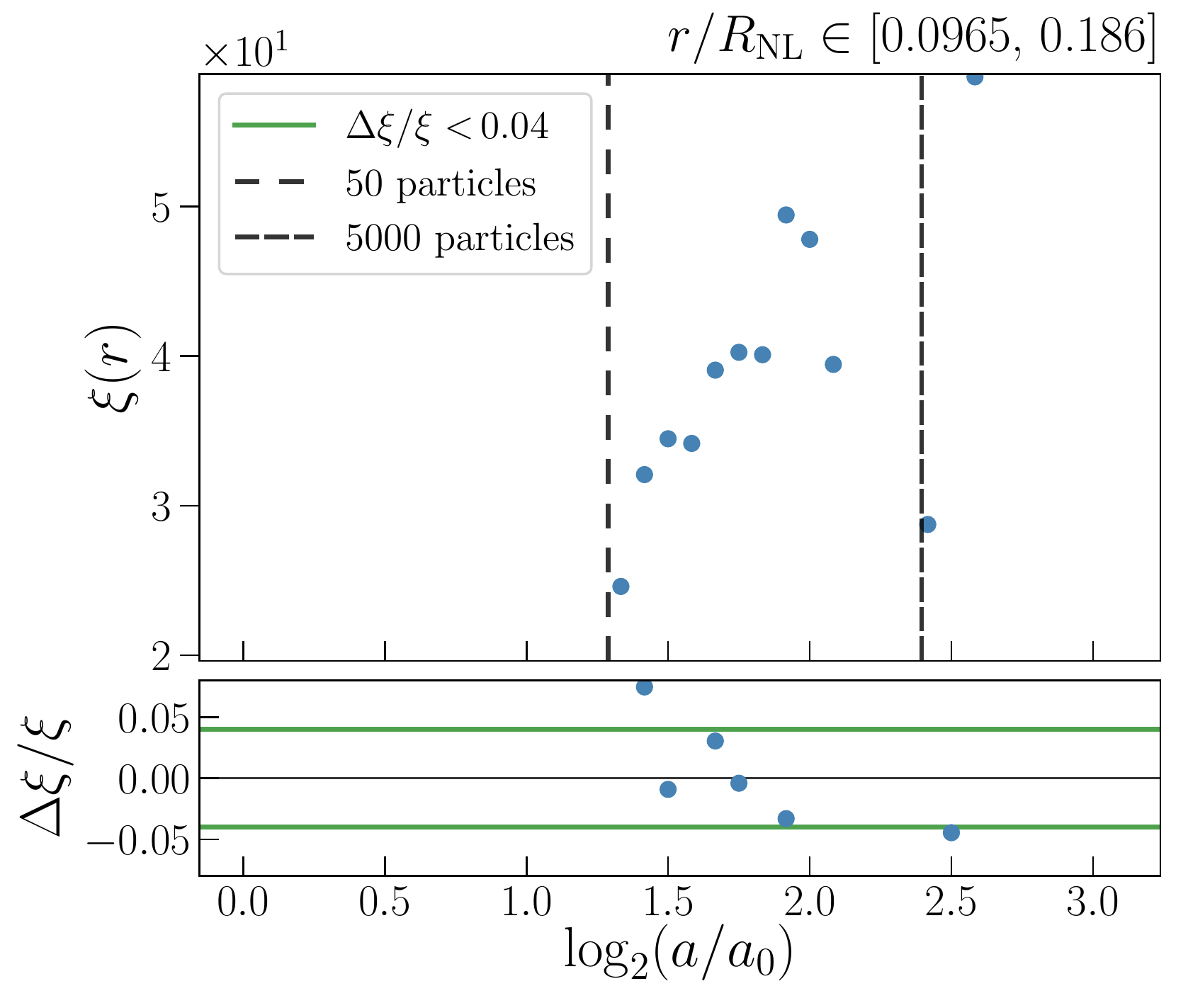}
    \end{subfigure}
    \hspace*{-0.7cm}
    \begin{subfigure}[b]{0.35\textwidth}
    \includegraphics[width=0.9\textwidth,height=0.85\linewidth]{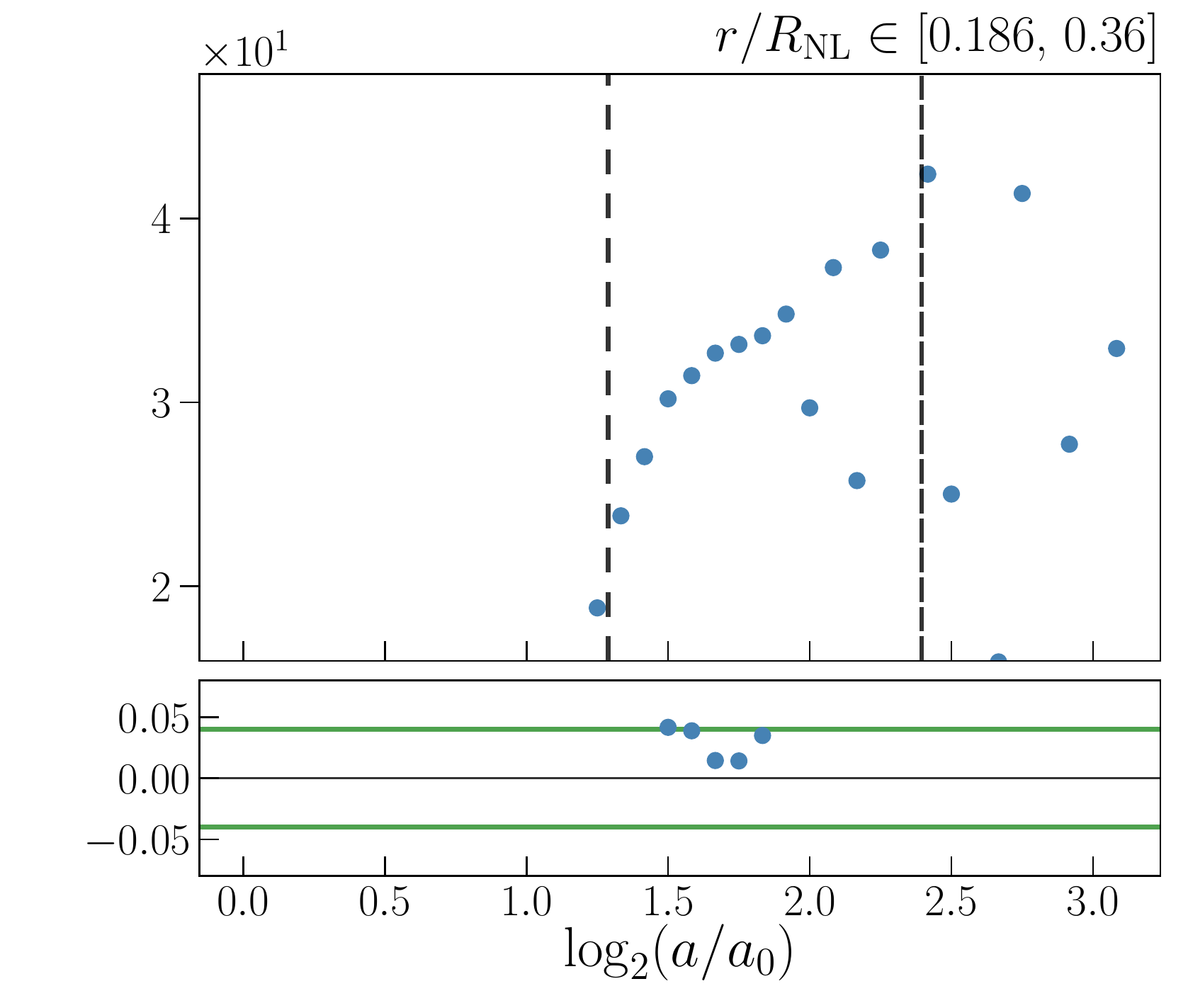}
    \end{subfigure}
    \hspace*{-0.7cm}
    \begin{subfigure}[b]{0.35\textwidth}
    \includegraphics[width=0.9\textwidth,height=0.85\linewidth]{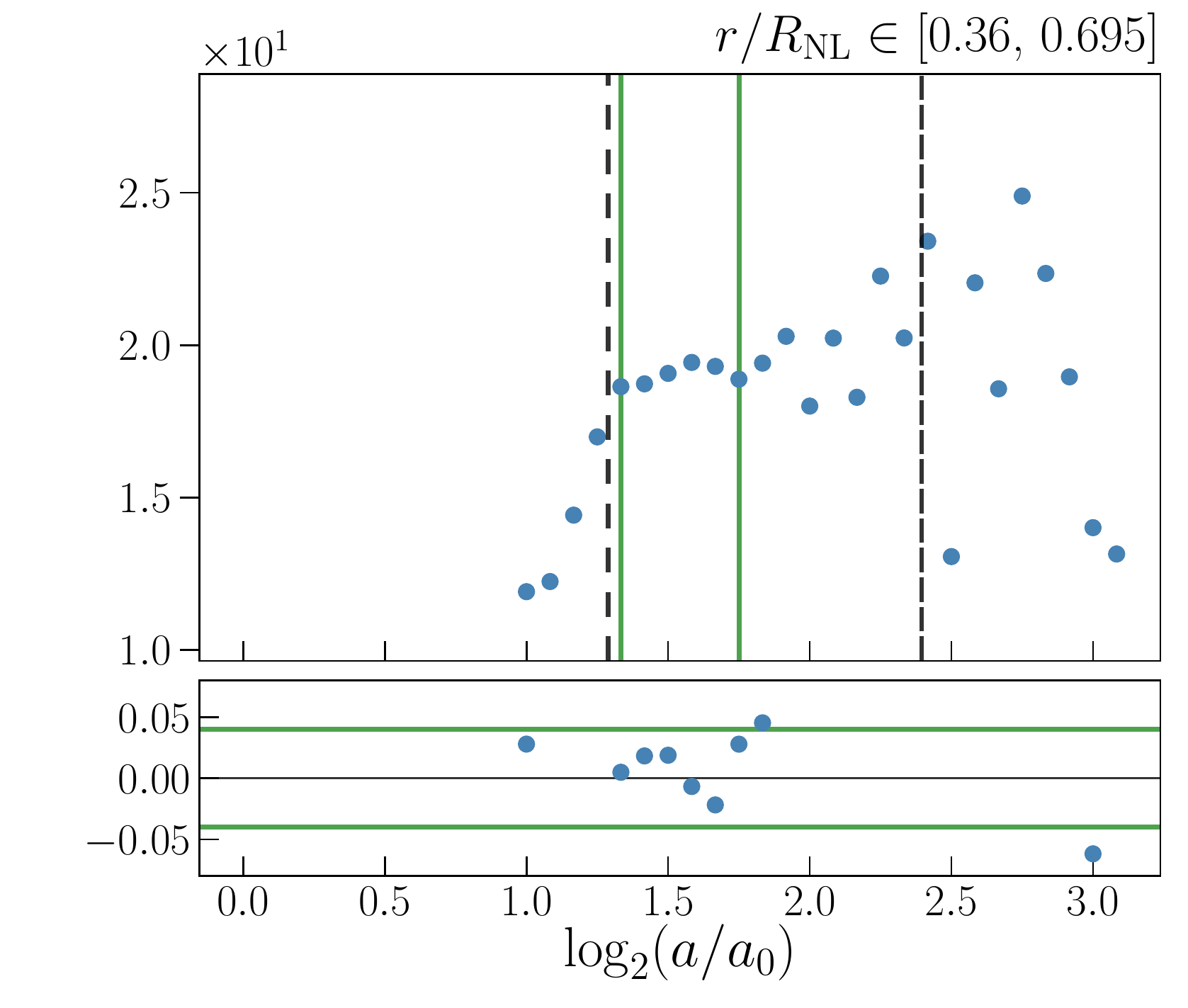}
    \end{subfigure}
    \\
    \centering
    \begin{subfigure}[b]{0.35\textwidth}
    \includegraphics[width=0.9\textwidth,height=0.85\linewidth]{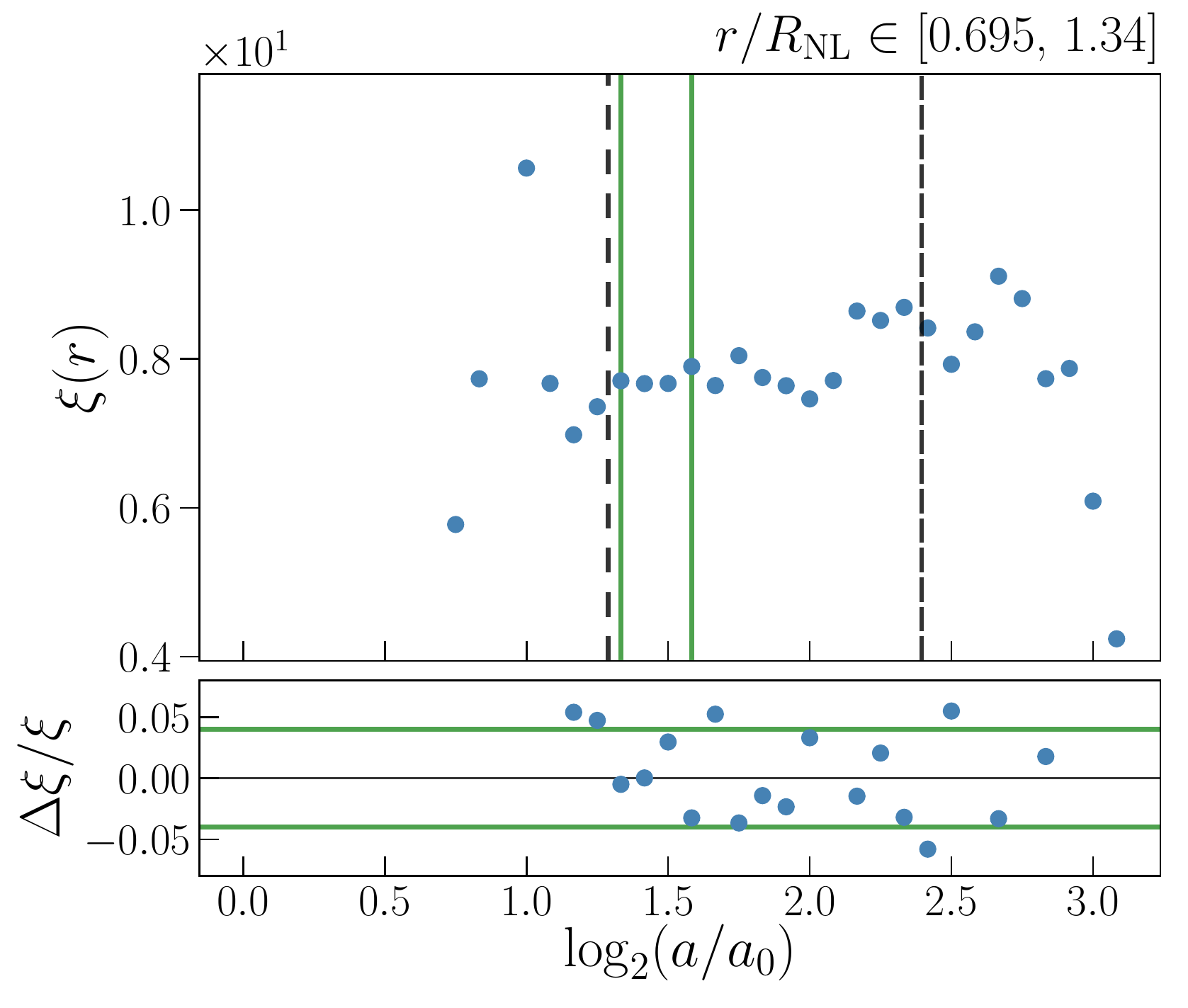}
    \end{subfigure}
    \hspace*{-0.7cm}
    \begin{subfigure}[b]{0.35\textwidth}
    \includegraphics[width=0.9\textwidth,height=0.85\linewidth]{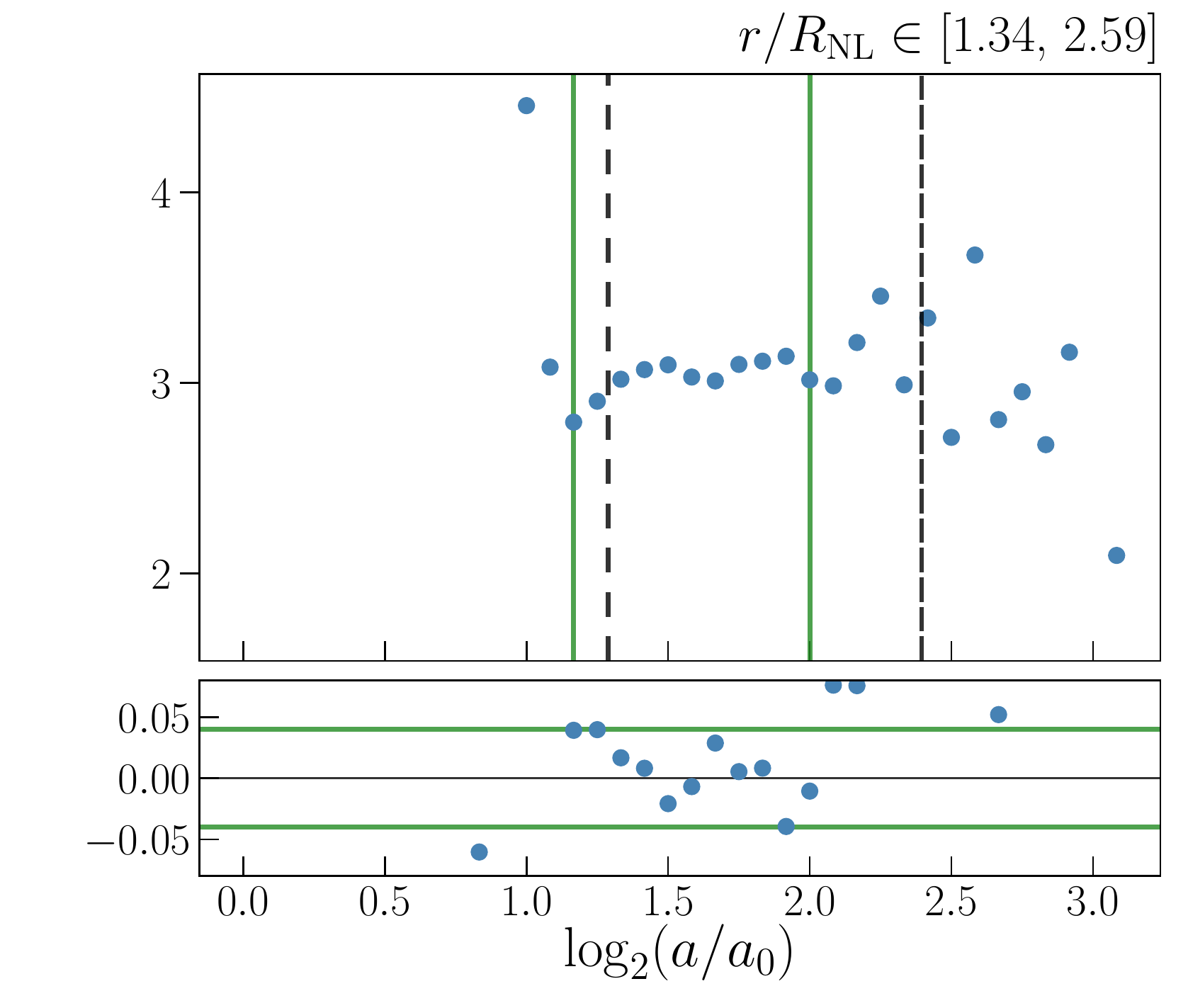}
    \end{subfigure} 
    \hspace*{-0.7cm}
    \begin{subfigure}[b]{0.35\textwidth}
    \includegraphics[width=0.9\textwidth,height=0.85\linewidth]{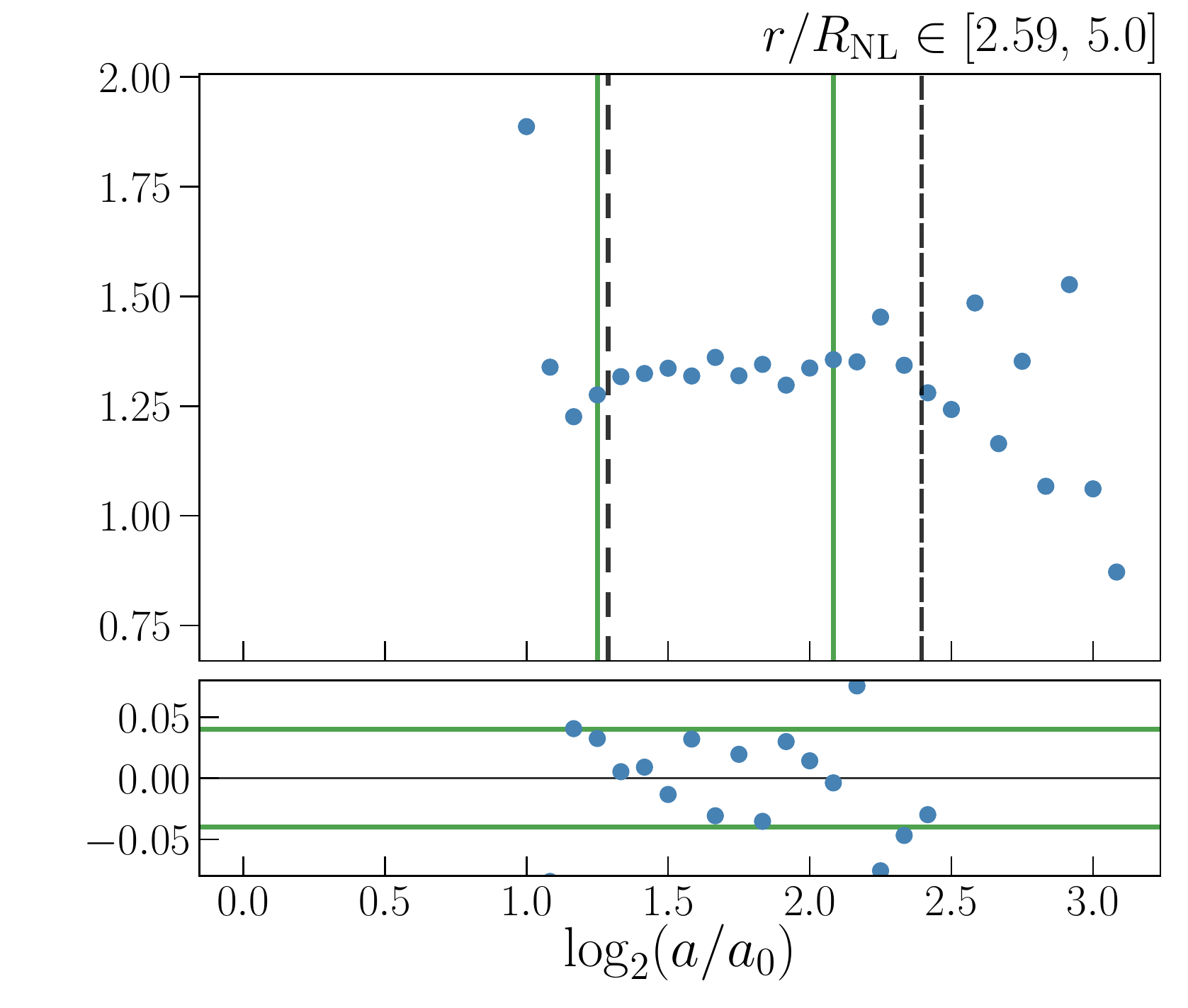}
    \end{subfigure}
    \\
\caption{Convergence of halo two-point correlation function for \Rockstar halos with rescaled mass $M/M_{\mathrm{NL}} \in [1.485,\,2.169]$. Each panel represents a bin of rescaled separation $r/R_{\mathrm{NL}}$ as indicated. As in Figure \ref{fig:ROCKSTAR_HCHC_CONVERGENCE}, the vertical green lines in the upper panels correspond to the most extended regions in which the fractional change $\Delta \xi / \xi$ is below $4\%$, also delimited by vertical lines in the lower panel. Note that virial radii of halos in this bin are  
$R_{\rm{vir}} \approx 0.2\,R_{\mathrm{NL}}$, so the first two to three bins probe the region in which halos overlap. 
}
\label{fig:ROCKSTAR_HCHC_CONVERGENCE}
\end{figure*}

\begin{figure}
\centering\resizebox{8cm}{!}{\includegraphics[]{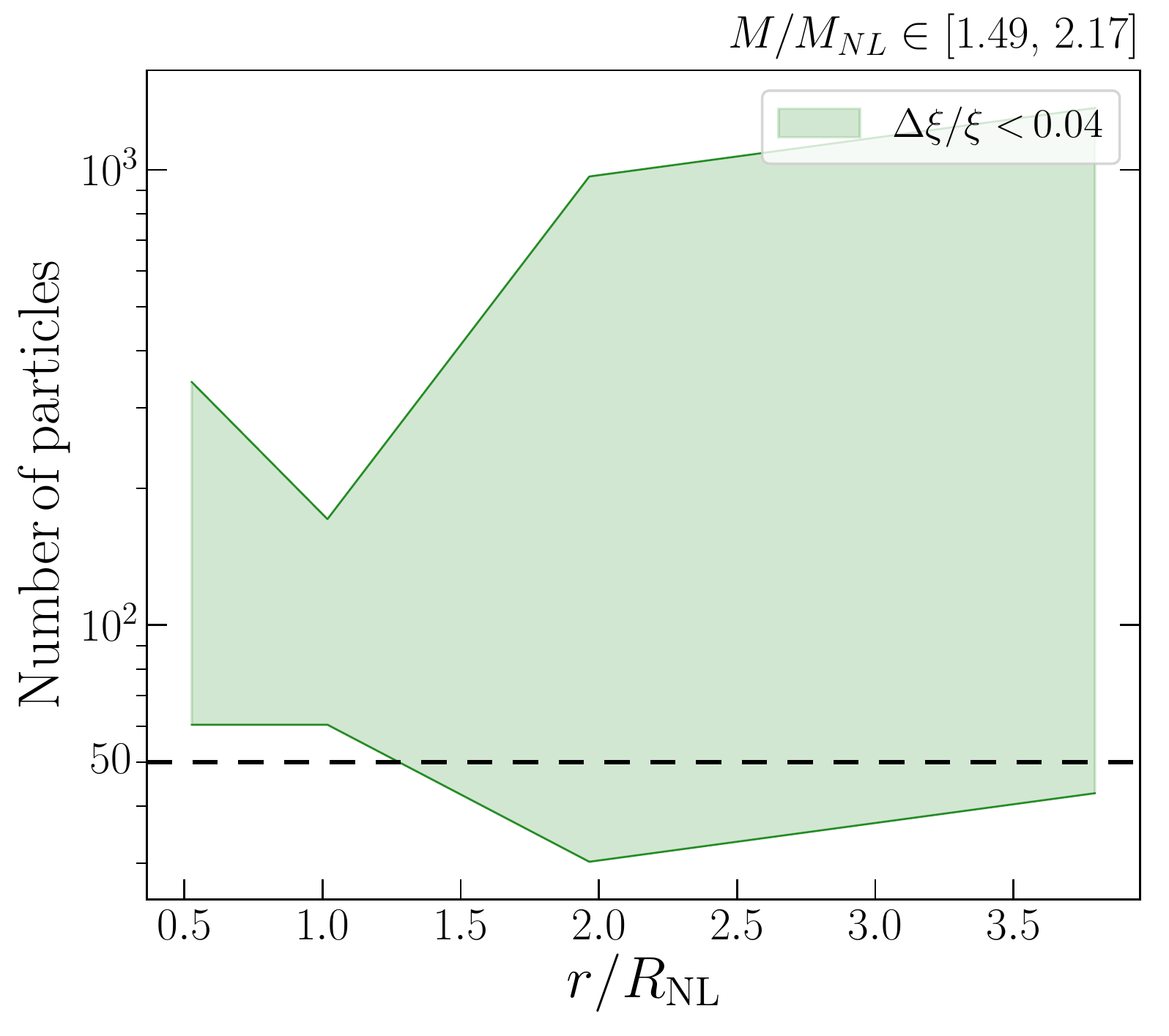}}
\caption{Mass resolution limits for halo 2PCF.  For halos in the same mass bin as Figure \ref{fig:ROCKSTAR_HCHC_CONVERGENCE}, $M/M_{\mathrm{NL}} \in [1.485,\,2.169]$,  the range (expressed in particle number per halo) in which convergence is observed, as a function of rescaled separations $r/R_{\mathrm NL}$ (corresponding to the centers of
the same bins as in Figure \ref{fig:ROCKSTAR_HCHC_CONVERGENCE}). The criteria used to
obtain these regions are analogous to those used for Figure \ref{fig:ROCKSTAR_HMF_RESOLUTION}, and are illustrated by the sets of green lines
in the panels if Figure \ref{fig:ROCKSTAR_HCHC_CONVERGENCE}). Note that this
plot is for the bins which satisfy the convergence criterion, and thus
excludes the first two bins in Figure \ref{fig:ROCKSTAR_HCHC_CONVERGENCE} in
which $\xi_{HH}$  appears to be systematically resolution dependent up to a limit of
at least several hundred particles per halo.}
\label{fig:ROCKSTAR_HCHC_CONVERGENCE_SCALE}
\end{figure}

We now consider the correlation properties of the clustering as characterised by $\xi_{HH}$, the two-point correlation function (2PCF) of halo centers. 
As discussed in the introduction, because $\xi_{HH}$ is in principal a function of both mass and separation, to test for  self-similarity 
we need to consider its value in bins of both rescaled mass and  
separation. Thus we divide the halos in bins of $M/M_{\mathrm{NL}}$ and
compare the measured 2PCF of the halos in a given bin, as a function of 
time, in bins of rescaled separation $r/R_{\mathrm{NL}}$.  We use again here the 
gravitationally bound SO virial mass and all halos (i.e. parent halos and subhalos).

As in P1 we have used the code {\it Corrfunc} \citep{2020MNRAS.491.3022S} to compute the 2PCFs. 

We do not present here results for the \FOF halos, 
because, as can be anticipated, the non-similarity of the HMF we have seen
leads to a strong breaking of self-similarity also in the halo 2PCF. Like for the HMF we find only at most some weak 
marginal convergence in $\xi_{HH}$, and our conclusion is that this finder is unsuitable for precision measurement of a physical clustering signal in halos, unless a lower cut-off of at least several thousand particles is used. To determine these bounds reliably would require considerably larger simulations than ours.

For \Rockstar halos our results are exemplified by those for the case displayed in Figure \ref{fig:ROCKSTAR_HCHC_CONVERGENCE}. As in our convergence analysis of the mass function, we use the catalog containing all halos (i.e. both parent halos and subhalos) labelled by their bound virial mass.
Figure \ref{fig:ROCKSTAR_HCHC_CONVERGENCE} shows, for a single chosen bin, $M/M_{\mathrm{NL}} \in [1.485,\,2.169]$, and in a similar format as used for the \Rockstar mass function,  the measured $\xi_{HH}$: each plot in Figure \ref{fig:ROCKSTAR_HCHC_CONVERGENCE} shows its evolution
as a function of time for a bin of rescaled separations with the limits indicated in the upper right hand of the plot. The mass binning is the same one from which the results shown in Figure \ref{fig:ROCKSTAR_HMF_CONVERGENCE}, and we have chosen this bin because it is in the intermediate range of mass where the best convergence of the mass functions is obtained. As in Figure \ref{fig:ROCKSTAR_HMF_CONVERGENCE} each plot has a subplot showing the fractional change $\Delta \xi/\xi$ in the 2PCF 
between consecutive snapshots.

While we observe no evidence for convergence in the first two bins, starting from the third we see a very similar behaviour to that for the mass functions:
above of order 50 particles there is apparent convergence, albeit with 
a residual level of fluctuations in the ``plateau" region that is significantly larger. In other words, at sufficiently large scales, we see
that $\xi_{HH}$ appears to converge well simply if 
the mass function is also well converged. Studying these same plots for other rescaled mass bins, and with different binnings, we find the same qualitative 
behaviours, and, for what concerns the lower bounds to convergence, the same
quantitative results. Further it becomes clear that the separation
scale characterising the observed transition to good convergence 
is simply set by the extent of the halos themselves. For the case plotted in the 
Figure \ref{fig:ROCKSTAR_HCHC_CONVERGENCE}, the  virial radii 
of halos in the mass bin are $R_{\rm{vir}} \approx 0.2\,R_{\mathrm{NL}}$, 
so the first two bins correspond to separations for which the 
halos overlap. Thus, in the catalog we are considering, which includes
all halos, these arise from contributions either from
parent-subhalo pairs or pairs of subhalos. In these bins 
there is clearly a measurable non-zero signal which remains
resolution dependent up to the thousands of particles per halo,
but the sampling noise (which increases for the 2PCF at small scales) 
is too large to allow us to determine robustly whether there is indeed convergence and at what resolution. Larger simulations would thus be required to
establish the resolution required to resolve the 2PCF at such small
scales. We note that the apparent marked problems with resolution at these scales is not in fact surprising, given the resolution issues we have discussed in the \FOF algorithm (which, as we have noted, is also employed in the initial identification of ``seed''  halos by
\Rockstar): \FOF applied to simulated halos will, because the particle density is finite,  include (exclude) regions below (above) the threshold density for percolation. The net effect  in cosmological halos (see e.g. \cite{2011ApJS..195....4M}) is an excess mass due to  spurious inclusion of mass in the halo's outskirts. As a consequence the mass assignment of very close-by seed halos can be biased, and unphysical correlations can result which may survive mass unbinding performed by \Rockstar. 

We proceed in the same manner as above for the mass function to state 
these results more precisely. Figure \ref{fig:ROCKSTAR_HCHC_CONVERGENCE_SCALE}
shows for halos with mass in the bin $M/M_{\mathrm{NL}} \in [1.485,\,2.169]$ 
the range of particle number per halo as 
a function of 
$r/R_{\mathrm{NL}}$ in which the simple convergence criterion illustrated by the green solid lines in Figure \ref{fig:ROCKSTAR_HCHC_CONVERGENCE} is satisfied.
The absence of a bound at small distances corresponds to the observed
lack of convergence, and we see the trend for convergence to improve strongly above a few times the typical virial radius in the mass bin.

\section{Conclusion}

Extending the methods applied in P1 to the full matter 2PCF, we have shown 
that self-similarity in scale-free cosmological $N$-body simulations provides a stringent test for the convergence to the physical (continuum) limit also of the statistical properties of halos extracted from such simulations. For halos selected using the simple \FOF algorithm, our analysis identifies and allows the precise quantification of a resolution dependence of the assigned mass 
which has been anticipated in previous work \citep{warren_et_al_2006,2011ApJS..195....4M} through study of isolated idealized halos. 
Analysing catalogs with SO virial masses output by the \Rockstar code, on the other hand, we have found halo mass functions which
converge well to a resolution independent value starting from of order 50 to 100 particles, and no measurable dependence 
on resolution for greater particle number at the level of precision we can
probe (conservatively, of order a few percent). For the halo-halo 2PCF (for \Rockstar)
we have found very similar results, albeit at a lower level of precision (conservatively, $5- 10\%$) and at separations 
above that at which contributing pairs can overlap. For smaller separations at which halos overlap we see, on the other hand, clear evidence for strong resolution dependence of the measured correlation amplitude up to at least several hundreds of particles per halo. 

Our results (like those in P1) have been derived for the $n=-2$ scale-free model
and (by construction) cannot be directly applied to physical cosmological models,
which are not scale-free. The essential general conclusions we have drawn are, nevertheless, stated in a model independent form, and even our more quantitative results can naively be extended to any model by simply identifying the relevant parameters 
(here, $M_{\mathrm NL}$ and $R_{\mathrm NL}$) with their appropriate (redshift dependent) values. Such a naive extrapolation assumes, however, that the results we have found do not depend on the difference between the $n=-2$ power spectrum and the physical one, are also unaffected by deviations from EdS cosmology. For what concerns the resolution effects we have quantified it appears very reasonable to assume that this is the case: these effects in principle, as we argued, arise from the halo finder and would not be expected to depend sensitively on the exact nature of the underlying density field on which they are used. Nevertheless in future work we will probe directly the model dependence of these results by performing the same analysis for different scale-free models. We will also explore further the dependence of our inferred constraints on other relevant parameters --- box size, force smoothing, parameters controlling initial conditions --- which we have not discussed at length here. Our results also show that it would also be potentially very instructive to perform significantly larger simulations of these models. As illustrated in particular by our results for the halo 2PCF at very small scales, by further reducing sampling noise, we should be able to determine the resolution necessary to obtain physically converged results at these scales rather than just diagnose that resolution is poor as we have done here.
Indeed similar considerations are very likely to apply for various correlation functions other than the single one we have considered here, and possibly more directly relevant to the construction of HOD models e.g. correlations between parent halos and subhalos or between halos in different mass ranges.

\section*{Acknowledgements} 
Both M.J. and M.L. warmly thank the Institute for Theory and Computation (ITC) for hosting them as, respectively, sabbatical visitor and internship student
during the academic year 2017-18 when this collaboration was initiated.
M.J. acknowledges in particular David Benhaiem and Benedikt Diemer for collaboration and many useful conversations on issues related to halos in self-similar cosmologies. We further thank Benedikt Diemer for providing us with data from his scale-free simulations which have been very useful for purposes of comparison, and also for useful comments on this article. D.J.E. is supported by the U.S. Department of Energy grant DE-SC0013178 and as a Simons Foundation Investigator. L.H.G. is also supported by the Simons Foundation. S.M. acknowledges the Fondation CFM pour la Recherche for financial support.

\section*{Data availability} 

The data reported in the article are provided in csv format in the online supplementary material.



\end{document}